\documentclass{aastex631}

\begin{document}

\title{No Glitch in the Matrix: Robust Reconstruction of Gravitational Wave Signals Under Noise Artifacts}

\author[0000-0001-8700-3455]{Chayan Chatterjee}
\affiliation{Department of Physics and Astronomy, Vanderbilt University\\ 2201 West End Avenue, Nashville, Tennessee - 37235,}
\affiliation{Data Science Institute, Vanderbilt University\\ 1400 18th Avenue South Building, Suite 2000, Nashville, Tennessee - 37212, }

\author[0000-0003-1007-8912]{Karan Jani}
\affiliation{Department of Physics and Astronomy, Vanderbilt University\\ 2201 West End Avenue, Nashville, Tennessee - 37235,} 



\begin{abstract}

Gravitational wave observations by ground-based detectors such as LIGO and Virgo have transformed astrophysics, enabling the study of compact binary systems and their mergers. However, transient noise artifacts, or ``glitches,” pose a significant challenge, often obscuring or mimicking signals and complicating their analysis. In this work, we extend the Attention-boosted Waveform Reconstruction network (\texttt{AWaRe}) to address glitch mitigation, demonstrating its robustness in reconstructing waveforms in the presence of real glitches from LIGO’s third observing run (O3). Without requiring explicit training on glitches, \texttt{AWaRe} accurately isolates gravitational wave signals from data contaminated by glitches spanning a wide range of amplitudes and morphologies. We evaluate this capability by investigating the events GW191109 and GW200129, which exhibit strong evidence of anti-aligned spins and spin-precession respectively, but may be adversely affected by data quality issues. We find that, regardless of the potential presence of glitches in the data, \texttt{AWaRe} reconstructs both waveforms with high accuracy. Additionally, we perform a systematic study of \texttt{AWaRe}'s performance on a simulated catalog of injected waveforms in real LIGO glitches and obtain reliable reconstructions of the waveforms. By subtracting the \texttt{AWaRe} reconstructions from the data, we show that the resulting residuals closely align with the background noise that the waveforms were injected in. The robustness of \texttt{AWaRe} in mitigating glitches, despite being trained exclusively on GW signals and not explicitly on glitches, highlights its potential as a powerful tool for improving the reliability of searches and characterizing noise artifacts.

\end{abstract}



\section{Introduction} \label{sec:intro}

Gravitational wave (GW) astronomy has made remarkable progress, with over 90 confirmed detections by the LIGO \citep{LIGO} and Virgo \citep{Virgo} observatories across three observing runs \citep{GWTC1, GWTC2, GWTC-2_1, GWTC-3}. These detections include a variety of astrophysical sources, such as binary black hole (BBH) mergers, binary neutron star (BNS) mergers, and neutron star-black hole (NSBH) mergers. The expanding catalog of events has provided valuable insights into the formation and evolution of compact binary systems, population-level properties of stellar remnants, and the fundamental physics governing these phenomena \citep{HubbleConstant, TGR, EOS, LVK_populations}. Additionally, detections like the BNS merger GW170817 \citep{GW170817}, accompanied by electromagnetic counterparts \citep{GW170817_1, GW170817_2, GW170817_EOS}, have established GWs as critical messengers for multi-messenger astronomy. \\

As detector sensitivities improve in upcoming observing runs, the number of detectable GW signals is expected to increase significantly \citep{GWTC-3}. However, the high frequency of transient noise artifacts, or ``glitches," in the data poses a major challenge \citep{GW170817_glitch, glitches_definition1, glitches_definition2, glitches_definition3}. Glitches, caused by various environmental and instrumental sources, can overlap with GW signals, obscuring or mimicking them \citep{Glitch_overlapping_with_signal}. With more signals likely to coincide with glitches in the future, these noise artifacts could severely impact the detection capabilities of GW observatories and the reliability of astrophysical inferences \citep{Impact_of_glitch1, Impact_of_glitch2, Impact_of_glitch3, Impact_of_glitch4, Impact_of_glitch5, Impact_of_glitch6}. Traditional glitch mitigation strategies, such as manual inspection and veto techniques, are labor-intensive and not scalable to the growing volume of data. Moreover, as new types of glitches continue to appear in detectors, it becomes increasingly important to develop model-independent approaches to glitch removal. Machine learning (ML) techniques, particularly neural networks, offer promising solutions by enabling automated, scalable, and generalizable methods for noise suppression and waveform reconstruction \citep{glitch1, glitch2, glitch3,glitch4, glitch5, glitch6, Chatterjee_2021, Bacon_denoising}. \\

In this work, we apply \texttt{AWaRe} or the Attention-boosted Waveform Reconstruction network  \citep{AWaRe, AWaRe_uncertainty} to recover GW waveforms from noisy detector data in the presence of glitches, without explicit training the model on glitch-augmented data. \texttt{AWaRe} is an encoder-decoder model consisting of a convolutional neural network (CNN) \citep{CNN_1, CNN_2} with attention mechanism \citep{Transformer}, and Long Short-Term Memory (LSTM) layers \citep{LSTM} trained on fully precessing BBH waveforms with higher-order modes. Previous studies have demonstrated that \texttt{AWaRe} can not only generate accurate point-estimate reconstructions of GW waveforms in the time-domain \citep{AWaRe}, but can also produce robust estimates of the reconstruction uncertainties \citep{AWaRe_uncertainty}, comparable to standard algorithms like BayesWave \citep{BayesWave}, coherent WaveBurst (cWB) \citep{cWB} and parameter estimation (PE) results. We show that \texttt{AWaRe}'s results are largely unaffected by glitches in the data by performing validation tests with simulated signals injected into real glitch data from O3 \citep{GWTC-3} and through extensive residual analysis. Furthermore, we study the two BBH events, GW191109 \citep{GW191109_1, GW191109_2, GWTC-3} and GW200129 \citep{GW200129_1, GW200129_2, GW200129_3, GW200129_4, GWTC-3}, and compare their reconstructions against published results.\\

This paper is organized as follows. In Section 2, we present the \texttt{AWaRe} reconstruction and residual analysis on synthetic data, consisting of simulated GW injections in real glitches from O3. In Section 3, we perform residual analysis after subtracting the \texttt{AWaRe} reconstructions from LIGO background data. In Section 4, we present the \texttt{AWaRe} reconstruction and residual analysis results for the O3 BBH events, GW191109 and GW200129 and compare the results with cWB and PE. Finally, in Section 5, we discuss the implications of this work and outline future directions. \\

\section{Reconstruction results for O3 glitches with overlapping GW injections} 

In this study, we conduct a systematic evaluation of the ability of \texttt{AWaRe} to reconstruct synthetic GW signals from data contaminated with real O3 glitches. For this analysis, we utilized the pre-trained \texttt{AWaRe} model described in \citep{AWaRe, AWaRe_uncertainty} directly, without any additional training or fine-tuning on datasets containing glitches. It is important to note that \texttt{AWaRe} was initially trained exclusively on GW strain data with no exposure to overlapping glitches. Despite this, the model had previously demonstrated robust performance and adaptability across various challenging scenarios beyond its training configuration \citep{AWaRe_uncertainty}. \\

The current investigation specifically aims to assess the generalizability of \texttt{AWaRe} under particularly difficult conditions—when glitches are present in the data alongside GW signals. By evaluating its behavior in such scenarios, we seek to understand how well the model retains its signal reconstruction accuracy, even when confronted with noise artifacts it was not explicitly designed to handle. This serves as a critical test of the model's robustness, highlighting its potential applicability to real-world GW data contaminated by non-Gaussian noise transients commonly observed in detectors. \\

\begin{figure}
\centering
\includegraphics[scale=0.7]{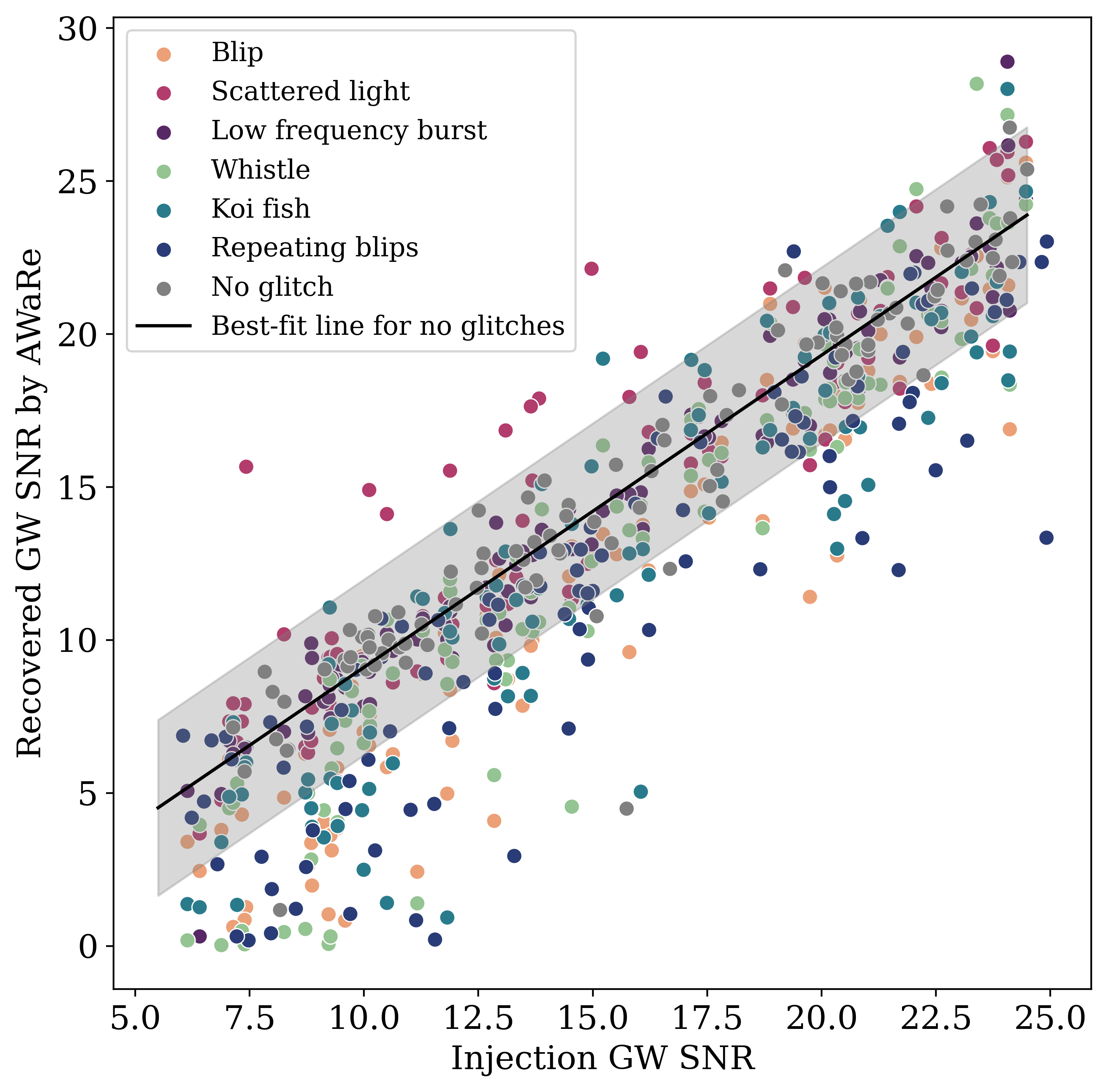}
\caption{Injected GW SNR (x-axis) and the recovered GW SNR (y-axis) for cases where GW signals overlap with glitches. The grey points represent cases without injected glitches, and the grey band shows the 90\% confidence interval of the recovered SNR in these no-glitch cases, assuming a Gaussian distribution. The black line indicates the best-fit line for the no-glitch data. The colored points indicate samples where GW signals overlap with different types of glitches.}
\label{fig:Figure_1}
\end{figure}

Fig.~\ref{fig:Figure_1} shows the distribution of SNRs of the injected GW signals and the reconstructed GW signals from \texttt{AWaRe} in the x and y axes respectively. To mimic realistic scenarios, the glitches were randomly selected from a catalog of real glitches observed in O3. These glitches were classified by GravitySpy \citep{GravitySpy, GravitySpy1, GravitySpy_classifier, GravitySpy_O3} with a confidence $>$ 0.9 and chosen to have SNRs in the same range as typically observed for real GW events in LIGO, allowing a direct investigation of the impact of glitches on reconstruction. We considered the following glitch classes in this study: Blips, Koi fish, Whistle, Low-frequency burst, Scattered light, Repeating blips. The segments of background O3 data around the glitches, used for all the analyses throughout this paper, was obtained from the Gravitational Wave Open Science Center (GWOSC) \citep{GWOSC}. The GW waveforms were generated using the IMRPhenomXPHM approximant \citep{IMRPhenomXPHM}, covering component masses between 10 and 80 M$_{\odot}$ in the detector frame, and spanning the full range of spin magnitudes and tilt angles. \\

The grey points in the figure show cases where no glitches were injected, representing the baseline reconstruction performance. The grey band denotes the 90\% confidence interval (CI) of the recovered SNR in these cases, assuming a Gaussian distribution, while the black line provides the best-fit line for no-glitch data. For samples without glitches, the reconstructed SNR aligns well with the injection SNR, with most points falling within the 90\% CI, demonstrating \texttt{AWaRe}'s accuracy in clean conditions. When glitches are injected, the recovered SNR shows increased scatter, particularly for certain glitch types such as Blips, Repeating Blips and Whistle. High-SNR glitches occasionally cause significant deviations from the no-glitch baseline. Additionally, more pronounced deviations are observed for lower SNR cases, as fainter signals are inherently more difficult to distinguish from noise and glitch artifacts. Despite these challenges, many glitch-injected samples remain close to the baseline, indicating \texttt{AWaRe}’s ability to effectively mitigate the effects of glitches in most scenarios. \\

\begin{figure}
\centering
\includegraphics[scale=0.47]{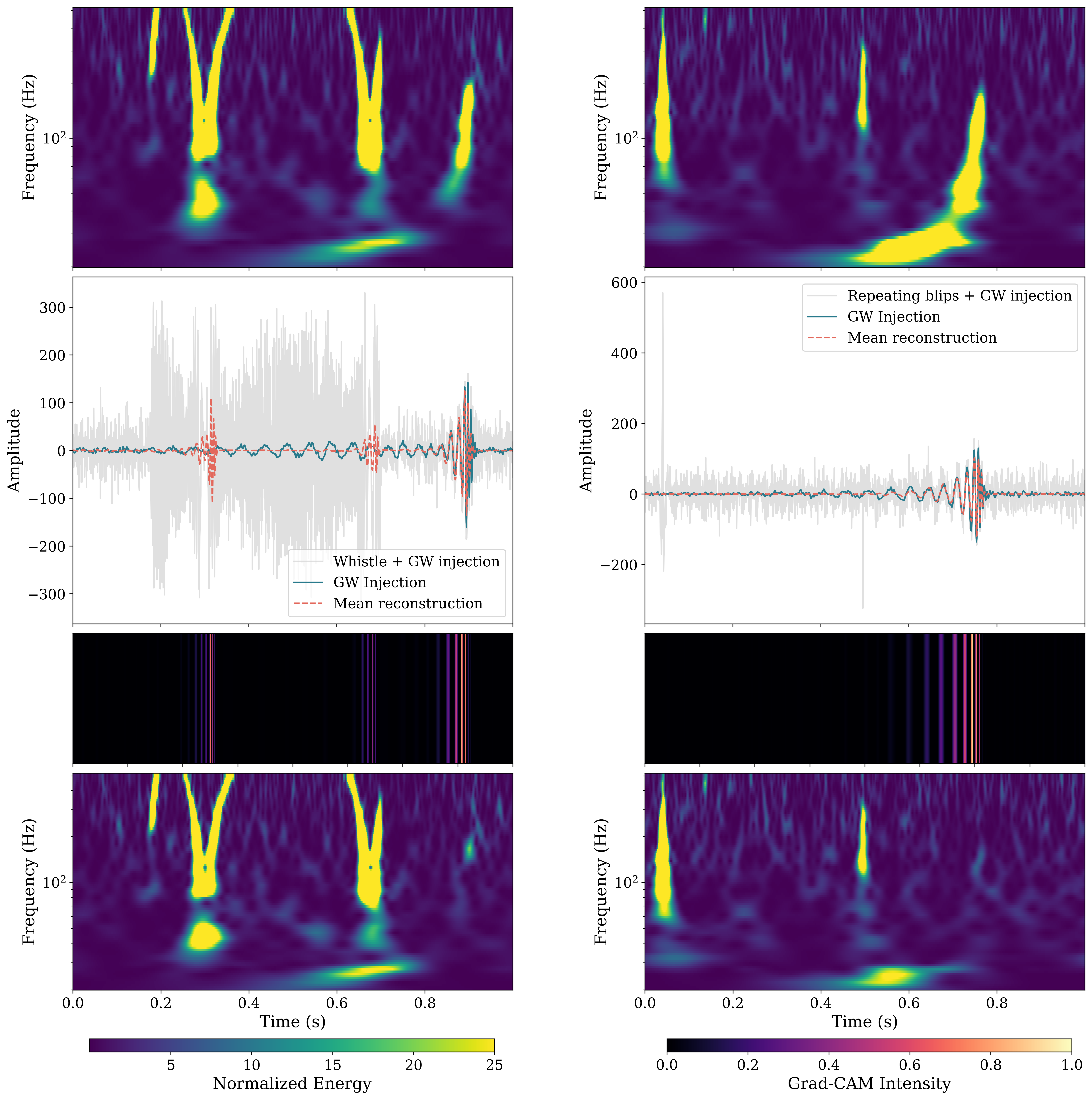}
\caption{Examples of reconstructions for simulated GW signals injected into O3 data segmentas containing glitches. The left panels correspond to a simulated GW signal overlaid with a whistle glitch, while the right panels show a GW signal with a repeating blip glitch. The top panels present the Q-transform of the input strain data, combining the simulated GW and the glitch. The second panels display the time-domain strain data (gray), the simulated GW waveform (blue), and the mean reconstruction from \texttt{AWaRe} (red dashed line). The third panels provide Grad-CAM visualizations, highlighting time regions critical to the model's reconstruction. The bottom panels show the residual Q-transform after subtracting the reconstructed waveform from the input strain data.}
\label{fig:Figure_2}
\end{figure}

Fig.~\ref{fig:Figure_2} shows examples of \texttt{AWaRe} reconstructions on this dataset. On the left, the input consists of a whistle glitch combined with a simulated GW signal, while the right shows a repeating blip glitch superimposed with a GW. The top panels depict the Q-transform of the combined strain data, illustrating the time-frequency features of the glitches and the GW chirps. For the whistle glitch (left), a narrow-band feature persists over a significant duration, overlapping with part of the GW chirp. In contrast, the repeating blip glitch (right) appears as a sequence of short, broadband bursts distributed across the signal duration. In both cases, the GW chirp remains discernible. \\

The second panels display the time-domain strain data, where the gray curve represents the combined input strain, the black line shows the injected GW signal, and the red dashed line represents the mean reconstruction from \texttt{AWaRe}. The left figure in the second panel shows an example of a scenario where the presence of a glitch adversely affects the reconstruction. Most of the GW signal overlapping with the glitch is not reconstructed because \texttt{AWaRe} is trained to output a vector of zeros when there is no GW in the data \citep{AWaRe, AWaRe_uncertainty}. Notably, at approximately 0.3 seconds and 0.7 seconds, the model outputs chirp-like features resembling a GW signal. This occurs because the glitch in these regions shows a rapid amplitude evolution from 0 to $\sim$ 300, a characteristic similar to the typical amplitude evolution of GW signals. As a result, the model misinterprets these regions as containing GW signals, demonstrating the influence of glitch characteristics on the reconstruction process. For the repeating blip glitch (right), the model performs better, reconstructing the GW signal with high accuracy, even in regions where the glitch overlaps. The robustness of the reconstruction highlights the model’s capability to focus on broadband GW features while ignoring transient, short-duration noise. \\

The third panels offer an interpretability tool through Grad-CAM visualizations \citep{Grad-CAM}. Grad-CAM (Gradient-weighted Class Activation Mapping) \citep{Grad-CAM} is a visualization technique used to interpret the decision-making process of deep learning models. It computes a weighted combination of features from the final layer of the network, guided by the gradient of the output with respect to these features.The Grad-CAM intensity, $L_{\text{Grad-CAM}}(t)$ at each time step $t$ is computed as,

\begin{equation}
    L_{\text{Grad-CAM}}(t) = \text{ReLU} \left( \sum_j \alpha_j \cdot w_j(t) \right)
\end{equation}

where $w_{j}(t)$ are the weights corresponding to the $j$-th neuron at time $t$ in the final dense layer, and $\alpha_{j}$ are the gradients of the model output $y$ with respect to the $j$-th weight, averaged over the time dimension:

\begin{equation}
    \alpha_j = \frac{1}{T} \sum_{t=1}^T \frac{\partial y}{\partial w_j(t)}
\end{equation}

Here, $T$ is the total number of time steps in the input. The ReLU function ensures that only positive contributions are visualized, highlighting the time steps most relevant for the reconstruction. The Grad-CAM visualizations highlight the time regions in the strain data where the model focuses to reconstruct the GW signal. For the whistle glitch, the Grad-CAM highlights are concentrated on the GW chirp, with reduced focus on the glitch-dominated regions, explaining the reconstruction challenges. For the repeating blip glitch, the activations align well with the GW chirp, demonstrating that the model successfully distinguishes the GW signal from the transient glitches. The bottom panels show the residual Q-transform obtained after subtracting the reconstructed waveform from the input strain data. For the whistle glitch, the residual Q-transform retains the narrow-band glitch. However, some residual signal power remains in the region overlapping the glitch, indicating incomplete reconstruction of the GW in this area. For the repeating blip glitch, the residual Q-transform shows that the glitch features persist, and only minimal residual signal power is visible, suggesting successful isolation and removal of the GW signal. 

\section{Analysis of Residuals} 

In this study, we have tested the accuracy of the residuals obtained by subtracting \texttt{AWaRe} reconstructions from data. For accurate GW reconstructions, the residuals should be consistent with the segment of background O3 data containing the glitch. To evaluate this, we used 100 examples of each glitch type: Blip, Koi fish, Whistle, Low-frequency burst, Scattered light, and Repeating blips. In each of these segments, we inject a GW waveform with component masses sampled between the ranges 10 and 80 M$_{\odot}$ and SNRs ranging between 6 to 25. We perform 10 such injections for each of the 100 glitches of a particular type. This gives us 1000 glitch + GW samples for each of the 6 types of glitches. We then obtain the GW reconstructions of all the 6000 samples using \texttt{AWaRe}. To obtain the residuals, we subtract the GW reconstructions from the glitch + GW samples: 

\begin{equation}
    \text{\texttt{AWaRe} residual = (GW waveform + O3 glitch data) – (\texttt{AWaRe} GW reconstruction).}
\label{eq:3}
\end{equation}

If the GW reconstructions are accurate, after subtracting the reconstructions from the data, the residuals should be similar to the segment of background O3 data around the glitch, before the GW waveform was injected. We verify this by subtracting the \texttt{AWaRe} residual from the original O3 glitch data and computing its optimal SNR,

\begin{equation}
    \text{SNR(Glitch data - \texttt{AWaRe} residual)}
\label{eq:4}
\end{equation}

If the \texttt{AWaRe} residual and the original glitch data match perfectly, this quantity should be 0. \\

\begin{figure}
\gridline{\fig{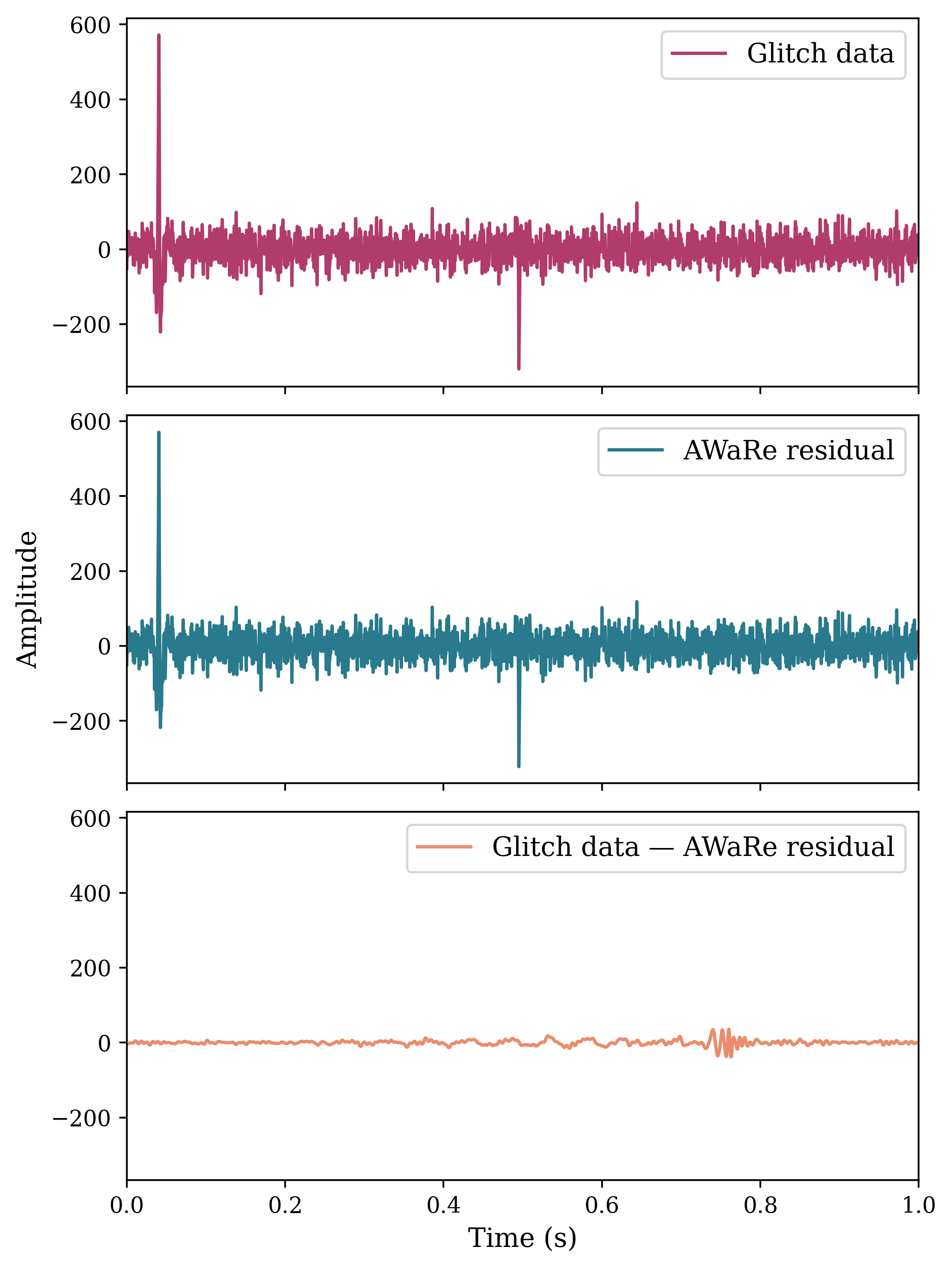}{0.36\textwidth}{(a)}
\fig{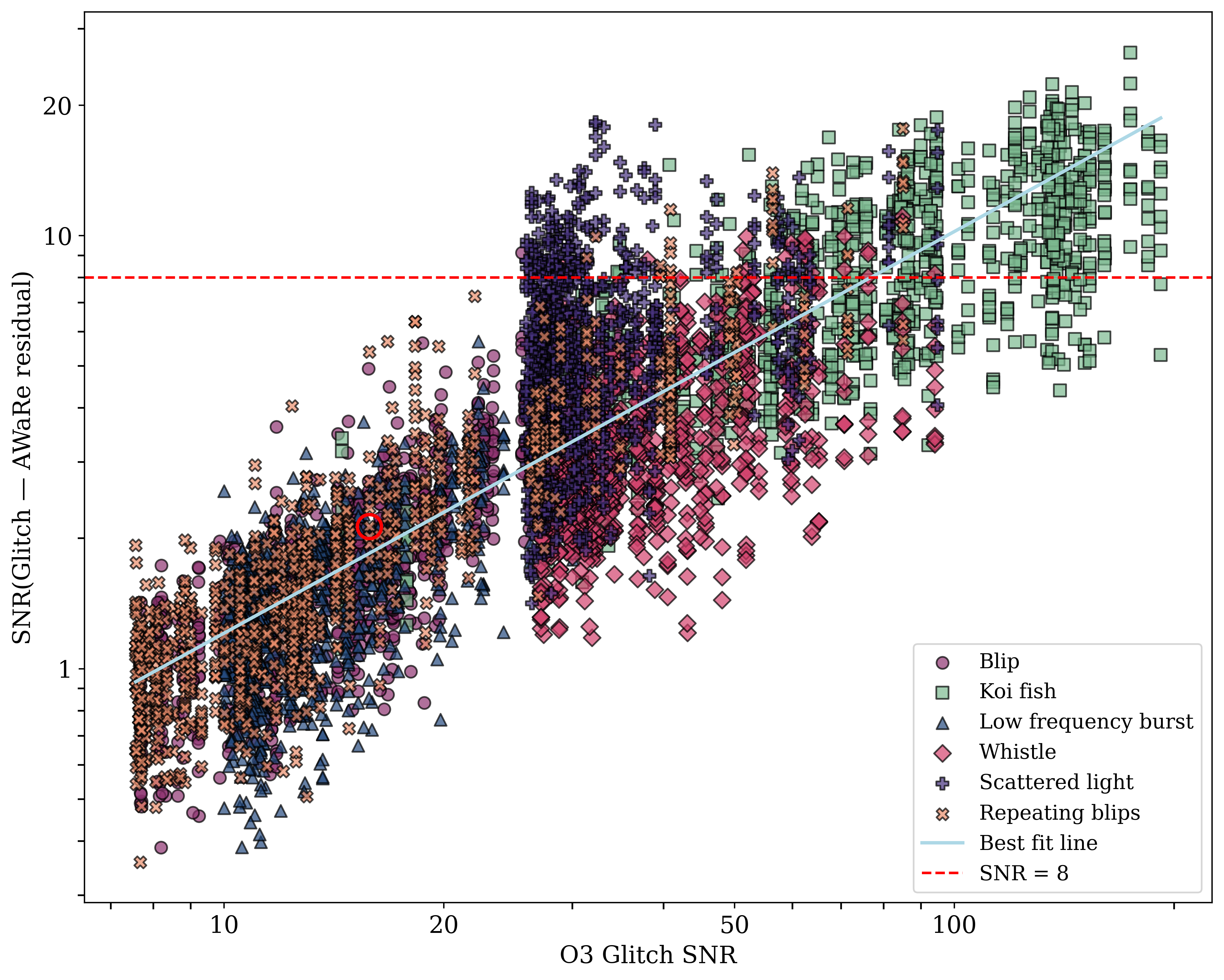}{0.6
  \textwidth}{(b)}
}
\gridline{\fig{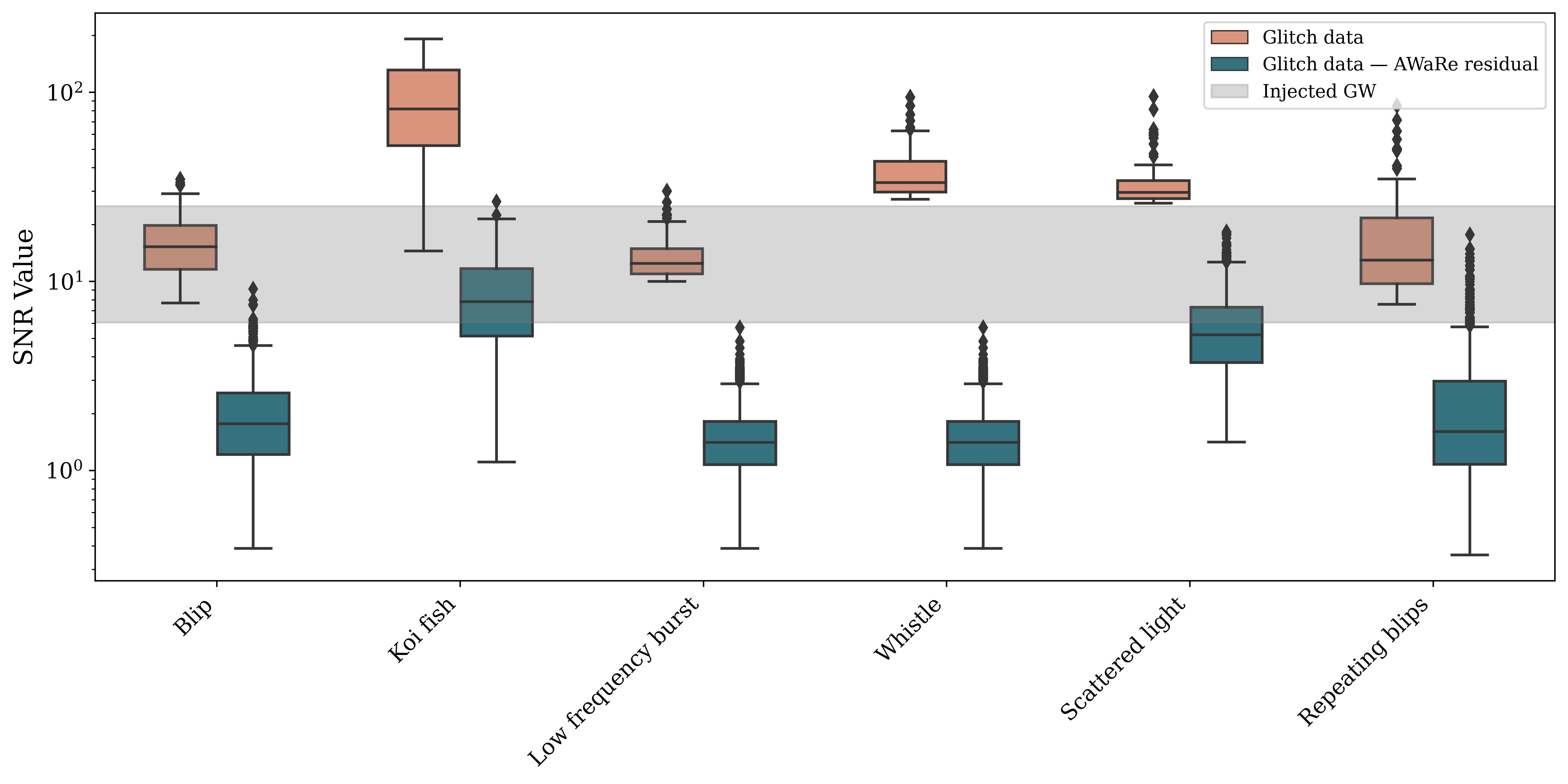}{0.9
  \textwidth}{(c)}}
\caption{\label{fig:Figure_3} (a) 1 sec O3 data around repeating blip glitch (top), \texttt{AWaRe} residual obtained by eq.~\ref{eq:3} (middle) and residual obtained by subtracting the curves in the top and middle panels (bottom) for the sample shown in Fig.~\ref{fig:Figure_2}. (b) Scatter plot showing the SNR values from eq.~\ref{eq:4}, plotted against the original glitch SNR, for various glitch types. Different colors and markers represent different glitch types. The blue line indicates the best-fit trend, showing a positive correlation between original glitch SNR and residual SNR. The red dashed line marks an SNR of 8, a common trigger threshold for online searches, included for reference. The red circle indicates the sample shown in (a). (c) Box and whisker plots showing the distributions of original glitch SNRs (orange) and the SNRs obtained from eq.~\ref{eq:4} (green). The gray band represents the range of SNRs for the GW signals injected into the data.}
\end{figure}

Fig.~\ref{fig:Figure_3} (a) shows the original glitch data (top), the \texttt{AWaRe} residual (middle) and the glitch - \texttt{AWaRe} residual data for the sample shown in Fig.~\ref{fig:Figure_2} (b). Since the GW reconstruction is accurate for this example, the residual and the original glitch data show significant similarity. Subtracting these two curves gives small fluctuations around zero amplitude around the region where the GW signal was injected, as seen in the bottom panel. \\ 

Fig.~\ref{fig:Figure_3} (b) displays the optimal SNR values of the (glitch - \texttt{AWaRe} residual) data as a function of the glitch SNRs. Each point in the figure represents a single sample, with different colors and shapes denoting different glitch types. For most samples, the SNR(glitch - \texttt{AWaRe} residual) values are significantly smaller than the original glitch SNR, indicating that our approach accurately reconstructs and removes the glitch in the majority of cases. The red dashed line marks the horizontal reference of SNR = 8, generally used as a trigger threshold in online searches, demonstrating that most residuals fall below this value, particularly for lower-SNR glitches. The scatter plot reveals a clear correlation between glitch SNR and residual SNR, with larger glitch SNRs resulting in larger residuals. This is highlighted by the best-fit line (blue), which shows a positive trend, indicating that while \texttt{AWaRe} performs well across a wide range of glitch strengths, higher-SNR glitches tend to leave behind stronger residuals. This trend suggests that the model's ability to accurately reconstruct glitches diminishes as the glitch amplitude increases. The example shown in Fig.~\ref{fig:Figure_3} (a) is indicated by the red circle. \\

Fig.~\ref{fig:Figure_3} (c) presents box and whisker plots summarizing the SNR distributions of original glitches and the (glitch - \texttt{AWaRe} residual) data. For most glitch types, such as Blips, Low-frequency bursts, and Repeating blips, the residual SNRs are substantially lower than the original glitch SNRs, demonstrating effective reconstruction. The gray band indicates the SNR range of the GW signals injected in the data. We notice that Koi fish glitches present challenges, as their residual SNRs often approach or even exceed the SNR threshold of 8, indicating incomplete reconstruction, especially for SNR $>$ 100. Overall, these figures highlight the robustness of \texttt{AWaRe} in reconstructing and removing glitches across a range of types and SNR values. \\

\section{Reconstruction results for GW191109 and GW200129}

GW191109 and GW200129 are significant BBH events that provide unique insights into binary formation and evolution \citep{GWTC-3}. GW191109, with its high masses and strong evidence for anti-aligned spin, is a compelling candidate for a dynamical formation scenario, potentially involving hierarchical mergers \citep{GW191109_1}. GW200129, the loudest BBH signal detected in O3, demonstrates evidence for spin-precession, highlighting the role of relativistic spin dynamics in black hole binary evolution \citep{GW200129_1, GW200129_2, GWTC-3}. Both events are affected by data quality issues that complicate their astrophysical interpretation \citep{GWTC-3}. For GW191109, a scattered light glitch in the LIGO Livingston data overlaps with key frequencies for spin inference, necessitating detailed analyses with multiple glitch models \citep{GWTC-3}. These tests reveal that anti-aligned spin inference is robust under some assumptions but can shift depending on the glitch model employed \citep{GW191109_2}. Similarly, for GW200129, a glitch in the LIGO Livingston detector coincides with the frequency range critical for precession inference. Re-analysis of the glitch-subtracted data and frequency-restricted studies indicate that the evidence for spin-precession depends sensitively on glitch modeling, underscoring the importance of rigorous data quality investigations \citep{GW200129_3, GW200129_4, GWTC-3, O3_glitch_subtraction}. \\

\begin{figure}
\gridline{\fig{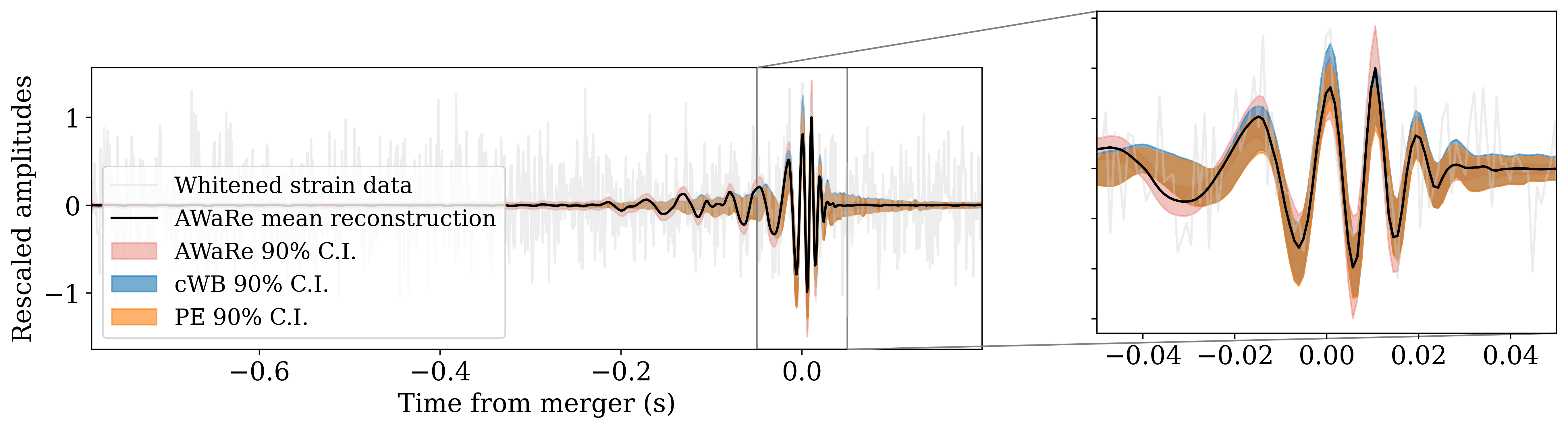}{0.9\textwidth}{(a)}}
\gridline{\fig{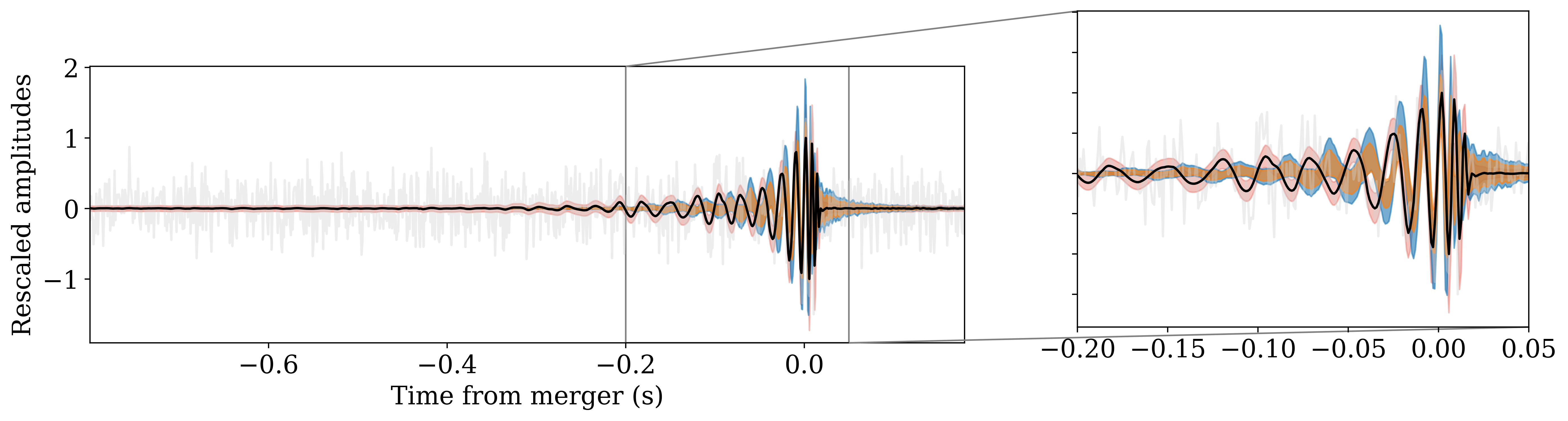}{0.9
  \textwidth}{(b)}}
\caption{\label{fig:Figure_4} \texttt{AWaRe} reconstruction results for (a) GW191109 and (b) GW200129. The gray curves show the whitened strain data. The black curves show the mean \texttt{AWaRe} reconstruction with the 90\% CI shown in red. The blue and orange bands show the 90\% CI from cWB and PE results respectively.}
\end{figure}

We perform waveform reconstruction using \texttt{AWaRe} to investigate these events. Fig.~\ref{fig:Figure_4} (a) and (b) shows the \texttt{AWaRe} mean reconstructions (black) and the 90\% CI (red), compared against the cWB (blue)\citep{cWB} and PE (orange) 90\% CI \citep{GWTC-3} for GW191109 and GW200109 respectively. The different analyses give consistent results around the merger. However, the \texttt{AWaRe} reconstruction for GW200129 shows mismatch in phase between 0.1 and 0.04 secs before merger, relative to cWB and PE results. \\

\begin{figure}
\centering
\includegraphics[scale=0.47]{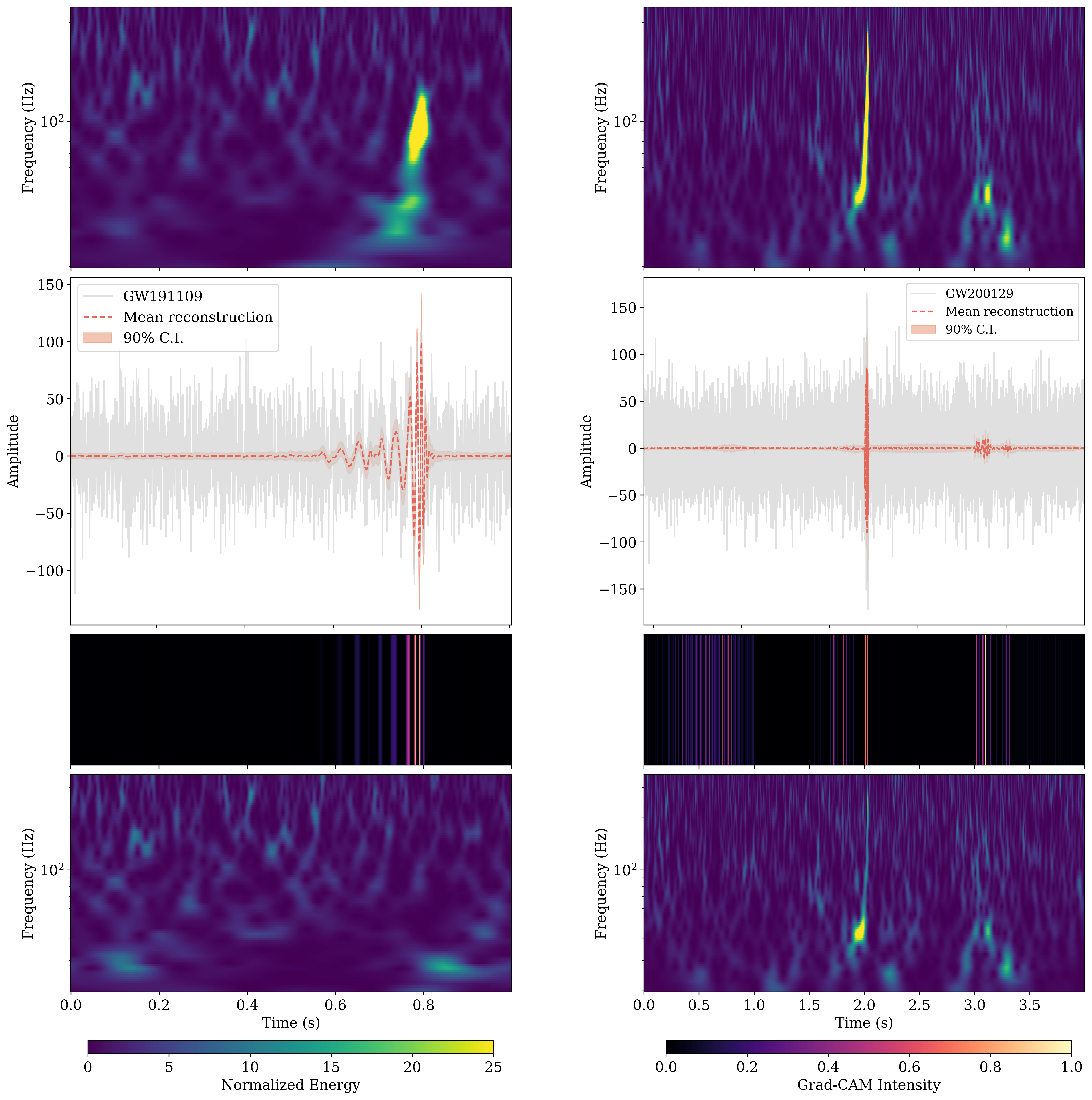}
\caption{Reconstruction results for GW191109 \citep{GWTC-3} (left) and GW200129 \citep{GWTC-3} (right) using \texttt{AWaRe}. The top panels display the Q-transform of the original strain data. The second panels show the time-domain strain (gray), the mean waveform reconstruction (red), and the associated 90\% confidence interval (shaded). The third panels present Grad-CAM visualizations, highlighting time regions critical for \texttt{AWaRe}'s reconstruction. The bottom panels illustrate the residual Q-transform obtained by subtracting the mean reconstruction from the strain data.}
\label{fig:Figure_5}
\end{figure}

Fig.~\ref{fig:Figure_5} demonstrates these results for GW191109 on the left and GW200129 on the right. The top panels display the Q-transform of the original strain data, providing a time-frequency representation of the power distribution. The second panels present the time-domain strain data (gray), the reconstructed waveform (red), and the associated 90\% confidence interval (shaded). For GW191109, the reconstruction aligns closely with the strain data, effectively capturing the morphology of the GW signal with high precision. For GW200129, the reconstruction captures much of the signal despite the strong presence of the glitch. Around the glitch at approximately 3 seconds, the output of \texttt{AWaRe} shows a noticeable increase in amplitude, indicating that the glitch is causing the model to trigger. However, the output amplitude is not high enough to significantly disrupt the model's ability to isolate the GW signal.  \\

The third panels present Grad-CAM visualizations, which identify the time regions where the model focuses to reconstruct the waveform. For GW191109, the highlighted regions closely align with the GW signal’s time of occurrence, demonstrating the model's precision in localizing and reconstructing the waveform. For GW200129, in addition to the merger region, the model activates near both the glitch and the inspiral, at 3 secs and between 0.5 and 1 sec respectively. However, while the output of the model is slightly elevated in the region of the glitch, the activations are not strong enough for the model to classify the glitch as part of a GW signal. \\

The bottom panels display the residual Q-transform obtained by subtracting the mean reconstructed waveform from the strain data. For GW191109, the residual plot shows no significant excess power, confirming that the model has effectively captured the GW. In contrast, for GW200129, the residual Q-transform reveals that the glitch remains largely intact after subtraction, with some residual signal power still present. Overall, these results emphasize the robustness of the \texttt{AWaRe} model in reconstructing GW signals from noisy data, even without explicit training on glitches, and its potential utility for GW data analysis in real scenarios. \\


\section{Discussion}

This work demonstrates the robustness of the \texttt{AWaRe} model in reconstructing GW signals amidst noise artifacts, as highlighted by our analysis of the GW events GW191109 and GW200129. Both events are notable for their astrophysical significance but are affected by data quality issues that complicate their interpretation. For GW191109, \texttt{AWaRe} successfully reconstructed the waveform with high accuracy, isolating the GW signal, leaving no significant excess power in the residual. For GW200129, which features a strong spin-precession signature, \texttt{AWaRe} effectively recovered the signal while leaving the glitch intact in the residuals. These results demonstrate the model’s ability to disentangle GWs even in the presence of overlapping noise artifacts. This capability addresses a critical challenge in GW astronomy, where transient noise artifacts frequently interfere with signal detection and analysis, particularly as detector sensitivities improve and new glitch types emerge. Beyond real events, we extended our analysis to a dataset of simulated GW signals injected into real glitches from O3. \texttt{AWaRe} consistently demonstrated robust performance across a wide range of glitch types, reconstructing the injected GWs with minimal contamination from overlapping noise. The systematic residual analysis further validates the model's ability to isolate noise artifacts, providing insights into the morphology and behavior of glitches. These findings establish \texttt{AWaRe} as a powerful, generalizable tool for GW data analysis that paves the way for more accurate astrophysical inferences.


\begin{acknowledgments}
The authors would like to thank Jess McIver and Derek Davis for helpful discussions on this work. This research was undertaken with the support of compute grant and resources, particularly the DGX A100 AI Computing Server, offered by the Vanderbilt Data Science Institute (DSI) located at Vanderbilt University, USA. This research used data obtained from the Gravitational Wave Open Science Center (https://www.gw-openscience.org), a service of LIGO Laboratory, the LIGO Scientific Collaboration and the Virgo Collaboration. LIGO is funded by the U.S. National Science Foundation. Virgo is funded by the French Centre National de Recherche Scientifique (CNRS), the Italian Istituto Nazionale della Fisica Nucleare (INFN) and the Dutch Nikhef, with contributions by Polish and Hungarian institutes. This material is based upon work supported by NSF's LIGO Laboratory which is a major facility fully funded by the National Science Foundation. 
\end{acknowledgments}

%

\vspace{5mm}






\bibliography{sample631}{}

\begin{thebibliography}{}
\expandafter\ifx\csname natexlab\endcsname\relax\def\natexlab#1{#1}\fi
\providecommand{\url}[1]{\href{#1}{#1}}
\providecommand{\dodoi}[1]{doi:~\href{http://doi.org/#1}{\nolinkurl{#1}}}
\providecommand{\doeprint}[1]{\href{http://ascl.net/#1}{\nolinkurl{http://ascl.net/#1}}}
\providecommand{\doarXiv}[1]{\href{https://arxiv.org/abs/#1}{\nolinkurl{https://arxiv.org/abs/#1}}}

\bibitem[{Aasi {et~al.}(2015)Aasi, Abbott, Abbott, Abbott, Abernathy, Ackley, Adams, Adams, Addesso, Adhikari, Adya, Affeldt, Aggarwal, Aguiar, Ain, Ajith, Alemic, Allen, Amariutei, Anderson, Anderson, \& Collaboration}]{LIGO}
Aasi, J., Abbott, B.~P., Abbott, R., {et~al.} 2015, Classical and Quantum Gravity, 32, 074001, \dodoi{10.1088/0264-9381/32/7/074001}

\bibitem[{Abbott {et~al.}(2016)Abbott, Abbott, Abbott, Abernathy, Acernese, Ackley, Adamo, Adams, Adams, Addesso, Adhikari, Adya, Affeldt, Agathos, Agatsuma, Aggarwal, Aguiar, Aiello, Ain, Ajith, Allen, Allocca, Altin, Anderson, Anderson, Arai, Araya, Arceneaux, Areeda, Arnaud, Arun, Ascenzi, Ashton, Ast, Aston, Astone, Aufmuth, Aulbert, Babak, Bacon, Bader, Baker, Baldaccini, Ballardin, Ballmer, Barayoga, Barclay, Barish, Barker, Barone, Barr, Barsotti, Barsuglia, Barta, Bartlett, Bartos, Bassiri, Basti, Batch, Baune, Bavigadda, Bazzan, Behnke, Bejger, Bell, Bell, Berger, Bergman, Bergmann, Berry, Bersanetti, Bertolini, Betzwieser, Bhagwat, Bhandare, Bilenko, Billingsley, Birch, Birney, Biscans, Bisht, Bitossi, Biwer, Bizouard, Blackburn, Blackburn, Blair, Blair, Blair, Bloemen, Bock, Bodiya, Boer, Bogaert, Bogan, Bohe, Bojtos, Bond, Bondu, Bonnand, Boom, Bork, Boschi, Bose, Bouffanais, Bozzi, Bradaschia, Brady, Braginsky, Branchesi, Brau, Briant, Brillet, Brinkmann, Brisson, Brockill, Brooks, Brown, Brown,
  Brown, Buchanan, Buikema, Bulik, Bulten, Buonanno, Buskulic, Buy, Byer, Cadonati, Cagnoli, Cahillane, Bustillo, Callister, Calloni, Camp, Cannon, Cao, Capano, Capocasa, Carbognani, Caride, Diaz, Casentini, Caudill, Cavaglià, Cavalier, Cavalieri, Cella, Cepeda, Baiardi, Cerretani, Cesarini, Chakraborty, Chalermsongsak, Chamberlin, Chan, Chao, Charlton, Chassande-Mottin, Chatterji, Chen, Chen, Cheng, Chincarini, Chiummo, Cho, Cho, Chow, Christensen, Chu, Chua, Chung, Ciani, Clara, Clark, Cleva, Coccia, Cohadon, Colla, Collette, Cominsky, Conte, Conti, Cook, Corbitt, Cornish, Corsi, Cortese, Costa, Coughlin, Coughlin, Coulon, Countryman, Couvares, Cowan, Coward, Cowart, Coyne, Coyne, Craig, Creighton, Cripe, Crowder, Cumming, Cunningham, Cuoco, Canton, Danilishin, D’Antonio, Danzmann, Darman, Dattilo, Dave, Daveloza, Davier, Davies, Daw, Day, DeBra, Debreczeni, Degallaix, Laurentis, Deléglise, Pozzo, Denker, Dent, Dereli, Dergachev, DeRosa, Rosa, DeSalvo, Dhurandhar, Díaz, Fiore, Giovanni, Lieto, Pace,
  Palma, Virgilio, Dojcinoski, Dolique, Donovan, Dooley, Doravari, Douglas, Downes, Drago, Drever, Driggers, Du, Ducrot, Dwyer, Edo, Edwards, Effler, Eggenstein, Ehrens, Eichholz, Eikenberry, Engels, Essick, Etzel, Evans, Evans, Everett, Factourovich, Fafone, Fair, Fairhurst, Fan, Fang, Farinon, Farr, Farr, Favata, Fays, Fehrmann, Fejer, Ferrante, Ferreira, Ferrini, Fidecaro, Fiori, Fiorucci, Fisher, Flaminio, Fletcher, Fournier, Franco, Frasca, Frasconi, Frei, Freise, Frey, Frey, Fricke, Fritschel, Frolov, Fulda, Fyffe, Gabbard, Gair, Gammaitoni, Gaonkar, Garufi, Gatto, Gaur, Gehrels, Gemme, Gendre, Genin, Gennai, George, Gergely, Germain, Ghosh, Ghosh, Giaime, Giardina, Giazotto, Gill, Glaefke, Goetz, Goetz, Gondan, González, Castro, Gopakumar, Gordon, Gorodetsky, Gossan, Gosselin, Gouaty, Graef, Graff, Granata, Grant, Gras, Gray, Greco, Green, Groot, Grote, Grunewald, Guidi, Guo, Gupta, Gupta, Gushwa, Gustafson, Gustafson, Hacker, Hall, Hall, Hammond, Haney, Hanke, Hanks, Hanna, Hannam, Hanson, Hardwick,
  Harms, Harry, Harry, Hart, Hartman, Haster, Haughian, Heidmann, Heintze, Heitmann, Hello, Hemming, Hendry, Heng, Hennig, Heptonstall, Heurs, Hild, Hoak, Hodge, Hofman, Hollitt, Holt, Holz, Hopkins, Hosken, Hough, Houston, Howell, Hu, Huang, Huerta, Huet, Hughey, Husa, Huttner, Huynh-Dinh, Idrisy, Indik, Ingram, Inta, Isa, Isac, Isi, Islas, Isogai, Iyer, Izumi, Jacqmin, Jang, Jani, Jaranowski, Jawahar, Jiménez-Forteza, Johnson, Jones, Jones, Jonker, Ju, K, Kalaghatgi, Kalogera, Kandhasamy, Kang, Kanner, Karki, Kasprzack, Katsavounidis, Katzman, Kaufer, Kaur, Kawabe, Kawazoe, Kéfélian, Kehl, Keitel, Kelley, Kells, Kennedy, Key, Khalaidovski, Khalili, Khan, Khan, Khan, Khazanov, Kijbunchoo, Kim, Kim, Kim, Kim, Kim, Kim, King, King, Kinzel, Kissel, Kleybolte, Klimenko, Koehlenbeck, Kokeyama, Koley, Kondrashov, Kontos, Korobko, Korth, Kowalska, Kozak, Kringel, Krishnan, Królak, Krueger, Kuehn, Kumar, Kuo, Kutynia, Lackey, Landry, Lange, Lantz, Lasky, Lazzarini, Lazzaro, Leaci, Leavey, Lebigot, Lee, Lee, Lee,
  Lee, Lenon, Leonardi, Leong, Leroy, Letendre, Levin, Levine, Li, Libson, Littenberg, Lockerbie, Logue, Lombardi, Lord, Lorenzini, Loriette, Lormand, Losurdo, Lough, Lück, Lundgren, Luo, Lynch, Ma, MacDonald, Machenschalk, MacInnis, Macleod, Magaña-Sandoval, Magee, Mageswaran, Majorana, Maksimovic, Malvezzi, Man, Mandel, Mandic, Mangano, Mansell, Manske, Mantovani, Marchesoni, Marion, Márka, Márka, Markosyan, Maros, Martelli, Martellini, Martin, Martin, Martynov, Marx, Mason, Masserot, Massinger, Masso-Reid, Matichard, Matone, Mavalvala, Mazumder, Mazzolo, McCarthy, McClelland, McCormick, McGuire, McIntyre, McIver, McManus, McWilliams, Meacher, Meadors, Meidam, Melatos, Mendell, Mendoza-Gandara, Mercer, Merilh, Merzougui, Meshkov, Messenger, Messick, Meyers, Mezzani, Miao, Michel, Middleton, Mikhailov, Milano, Miller, Millhouse, Minenkov, Ming, Mirshekari, Mishra, Mitra, Mitrofanov, Mitselmakher, Mittleman, Moggi, Mohan, Mohapatra, Montani, Moore, Moore, Moraru, Moreno, Morriss, Mossavi, Mours,
  Mow-Lowry, Mueller, Mueller, Muir, Mukherjee, Mukherjee, Mukherjee, Mukund, Mullavey, Munch, Murphy, Murray, Mytidis, Nardecchia, Naticchioni, Nayak, Necula, Nedkova, Nelemans, Neri, Neunzert, Newton, Nguyen, Nielsen, Nissanke, Nitz, Nocera, Nolting, Normandin, Nuttall, Oberling, Ochsner, O’Dell, Oelker, Ogin, Oh, Oh, Ohme, Oliver, Oppermann, Oram, O’Reilly, O’Shaughnessy, Ottaway, Ottens, Overmier, Owen, Pai, Pai, Palamos, Palashov, Palomba, Pal-Singh, Pan, Pankow, Pannarale, Pant, Paoletti, Paoli, Papa, Paris, Parker, Pascucci, Pasqualetti, Passaquieti, Passuello, Patricelli, Patrick, Pearlstone, Pedraza, Pedurand, Pekowsky, Pele, Penn, Perreca, Phelps, Piccinni, Pichot, Piergiovanni, Pierro, Pillant, Pinard, Pinto, Pitkin, Poggiani, Popolizio, Post, Powell, Prasad, Predoi, Premachandra, Prestegard, Price, Prijatelj, Principe, Privitera, Prodi, Prokhorov, Puncken, Punturo, Puppo, Pürrer, Qi, Qin, Quetschke, Quintero, Quitzow-James, Raab, Rabeling, Radkins, Raffai, Raja, Rakhmanov, Rapagnani,
  Raymond, Razzano, Re, Read, Reed, Regimbau, Rei, Reid, Reitze, Rew, Reyes, Ricci, Riles, Robertson, Robie, Robinet, Rocchi, Rolland, Rollins, Roma, Romano, Romanov, Romie, Rosińska, Rowan, Rüdiger, Ruggi, Ryan, Sachdev, Sadecki, Sadeghian, Salconi, Saleem, Salemi, Samajdar, Sammut, Sanchez, Sandberg, Sandeen, Sanders, Sassolas, Sathyaprakash, Saulson, Sauter, Savage, Sawadsky, Schale, Schilling, Schmidt, Schmidt, Schnabel, Schofield, Schönbeck, Schreiber, Schuette, Schutz, Scott, Scott, Sellers, Sengupta, Sentenac, Sequino, Sergeev, Serna, Setyawati, Sevigny, Shaddock, Shah, Shahriar, Shaltev, Shao, Shapiro, Shawhan, Sheperd, Shoemaker, Shoemaker, Siellez, Siemens, Sigg, Silva, Simakov, Singer, Singer, Singh, Singh, Singhal, Sintes, Slagmolen, Slutsky, Smith, Smith, Smith, Son, Sorazu, Sorrentino, Souradeep, Srivastava, Staley, Steinke, Steinlechner, Steinlechner, Steinmeyer, Stephens, Stone, Strain, Straniero, Stratta, Strauss, Strigin, Sturani, Stuver, Summerscales, Sun, Sutton, Swinkels,
  Szczepańczyk, Tacca, Talukder, Tanner, Tápai, Tarabrin, Taracchini, Taylor, Theeg, Thirugnanasambandam, Thomas, Thomas, Thomas, Thorne, Thorne, Thrane, Tiwari, Tiwari, Tokmakov, Tomlinson, Tonelli, Torres, Torrie, Töyrä, Travasso, Traylor, Trifirò, Tringali, Trozzo, Tse, Turconi, Tuyenbayev, Ugolini, Unnikrishnan, Urban, Usman, Vahlbruch, Vajente, Valdes, van Bakel, van Beuzekom, van~den Brand, Broeck, Vander-Hyde, van~der Schaaf, van Heijningen, van Veggel, Vardaro, Vass, Vasúth, Vaulin, Vecchio, Vedovato, Veitch, Veitch, Venkateswara, Verkindt, Vetrano, Viceré, Vinciguerra, Vine, Vinet, Vitale, Vo, Vocca, Vorvick, Voss, Vousden, Vyatchanin, Wade, Wade, Wade, Walker, Wallace, Walsh, Wang, Wang, Wang, Wang, Wang, Ward, Warner, Was, Weaver, Wei, Weinert, Weinstein, Weiss, Welborn, Wen, Weßels, Westphal, Wette, Whelan, Whitcomb, White, Whiting, Williams, Williamson, Willis, Willke, Wimmer, Winkler, Wipf, Wittel, Woan, Worden, Wright, Wu, Yablon, Yam, Yamamoto, Yancey, Yap, Yu, Yvert, Zadrożny,
  Zangrando, Zanolin, Zendri, Zevin, Zhang, Zhang, Zhang, Zhang, Zhao, Zhou, Zhou, Zhu, Zotov, Zucker, Zuraw, Zweizig, Collaboration, \& Collaboration)}]{glitches_definition1}
Abbott, B.~P., Abbott, R., Abbott, T.~D., {et~al.} 2016, Classical and Quantum Gravity, 33, 134001, \dodoi{10.1088/0264-9381/33/13/134001}

\bibitem[{Abbott {et~al.}(2017{\natexlab{a}})Abbott, Abbott, Abbott, Acernese, Ackley, Adams, Adams, Addesso, Adhikari, Adya, Affeldt, Afrough, Agarwal, Agathos, Collaboration, \& Collaboration}]{HubbleConstant}
---. 2017{\natexlab{a}}, Nature, 551, 85, \dodoi{10.1038/nature24471}

\bibitem[{Abbott {et~al.}(2017{\natexlab{b}})Abbott, Abbott, Abbott, Acernese, Ackley, Adams, Adams, Addesso, Adhikari, Adya, Affeldt, Afrough, Agarwal, Agathos, Agatsuma, Aggarwal, Aguiar, Aiello, Ain, Ajith, Allen, Allen, Allocca, Altin, Amato, Ananyeva, Anderson, \& Anderson}]{GW170817}
---. 2017{\natexlab{b}}, Phys. Rev. Lett., 119, 161101, \dodoi{10.1103/PhysRevLett.119.161101}

\bibitem[{Abbott {et~al.}(2017{\natexlab{c}})Abbott, Abbott, Abbott, Acernese, Ackley, Adams, Adams, Addesso, Adhikari, Adya, Affeldt, Afrough, Agarwal, Agathos, Agatsuma, Aggarwal, Aguiar, Aiello, Ain, Ajith, Allen, Allen, Allocca, Aloy, \& Altin}]{GW170817_1}
---. 2017{\natexlab{c}}, The Astrophysical Journal Letters, 848, L12, \dodoi{10.3847/2041-8213/aa91c9}

\bibitem[{Abbott {et~al.}(2017{\natexlab{d}})Abbott, Abbott, Abbott, Acernese, Ackley, Adams, Adams, Addesso, Adhikari, Adya, Affeldt, Afrough, Agarwal, Agathos, Agatsuma, Aggarwal, Aguiar, Aiello, Ain, Ajith, Allen, Allen, Allocca, Aloy, \& Altin}]{GW170817_2}
---. 2017{\natexlab{d}}, The Astrophysical Journal Letters, 848, L13, \dodoi{10.3847/2041-8213/aa920c}

\bibitem[{Abbott {et~al.}(2019{\natexlab{a}})Abbott, Abbott, Abbott, Abraham, Acernese, Ackley, Adams, Adhikari, Adya, Affeldt, Agathos, Agatsuma, Aggarwal, Aguiar, Aiello, Ain, Ajith, Allen, Allocca, Aloy, Altin, Amato, Ananyeva, Anderson, Anderson, Angelova, Antier, Appert, Arai, Araya, Areeda, Ar\`ene, Arnaud, \& Arun}]{GWTC1}
---. 2019{\natexlab{a}}, Phys. Rev. X, 9, 031040, \dodoi{10.1103/PhysRevX.9.031040}

\bibitem[{Abbott {et~al.}(2019{\natexlab{b}})Abbott, Abbott, Abbott, Acernese, Ackley, Adams, Adams, Addesso, Adhikari, Adya, Affeldt, Agarwal, Agathos, Agatsuma, Aggarwal, Aguiar, Aiello, Ain, Ajith, Allen, Allen, Allocca, Aloy, Altin, Amato, Ananyeva, Anderson, Anderson, Angelova, Antier, Appert, Arai, Araya, \& Areeda}]{EOS}
---. 2019{\natexlab{b}}, Phys. Rev. X, 9, 011001, \dodoi{10.1103/PhysRevX.9.011001}

\bibitem[{Abbott {et~al.}(2019{\natexlab{c}})Abbott, Abbott, Abbott, Abraham, Acernese, Ackley, Adams, Adhikari, Adya, Affeldt, Agathos, Agatsuma, Aggarwal, Aguiar, Aiello, Ain, Ajith, Allen, Allocca, Aloy, Altin, Amato, Ananyeva, Anderson, Anderson, Angelova, Antier, Appert, Arai, Araya, Areeda, Arène, Arnaud, Arun, Ascenzi, Ashton, Aston, Astone, Aubin, Aufmuth, AultONeal, Austin, Avendano, Avila-Alvarez, Babak, Collaboration, \& the Virgo~Collaboration}]{LVK_populations}
---. 2019{\natexlab{c}}, The Astrophysical Journal Letters, 882, L24, \dodoi{10.3847/2041-8213/ab3800}

\bibitem[{Abbott {et~al.}(2021)Abbott, Abbott, Abraham, Acernese, Ackley, Adams, Adams, Adhikari, Adya, Affeldt, Agathos, Agatsuma, Aggarwal, Aguiar, Aiello, Ain, Ajith, Akcay, Allen, Allocca, Altin, Amato, Anand, Ananyeva, Anderson, Anderson, \& Angelova}]{GWTC2}
Abbott, R., Abbott, T.~D., Abraham, S., {et~al.} 2021, Phys. Rev. X, 11, 021053, \dodoi{10.1103/PhysRevX.11.021053}

\bibitem[{Abbott {et~al.}(2023)Abbott, Abbott, Acernese, Ackley, Adams, Adhikari, Adhikari, Adya, Affeldt, Agarwal, Agathos, Agatsuma, Aggarwal, Aguiar, Aiello, Ain, Ajith, Akcay, Akutsu, Albanesi, Allocca, Altin, Amato, Anand, Anand, Ananyeva, Anderson, Anderson, Ando, Andrade, Andres, Angelova, Ansoldi, Antelis, Antier, Appert, Arai, Arai, Arai, Araki, Araya, Araya, Areeda, Ar\`ene, Aritomi, Arnaud, Arogeti, Aronson, Arun, Asada, Asali, Ashton, Aso, Assiduo, Aston, Astone, Aubin, Austin, Babak, Badaracco, Bader, Badger, Bae, Bae, Baer, Bagnasco, Bai, Baiotti, Baird, Bajpai, Ball, Ballardin, Ballmer, Balsamo, \& Baltus}]{GWTC-3}
Abbott, R., Abbott, T.~D., Acernese, F., {et~al.} 2023, Phys. Rev. X, 13, 041039, \dodoi{10.1103/PhysRevX.13.041039}

\bibitem[{Abbott {et~al.}(2024)Abbott, Abbott, Acernese, Ackley, Adams, Adhikari, Adhikari, Adya, Affeldt, Agarwal, Agathos, Agatsuma, Aggarwal, Aguiar, Aiello, Ain, Ajith, Albanesi, Allocca, Altin, Amato, Anand, Anand, Ananyeva, Anderson, Anderson, Andrade, Andres, Andri\ifmmode~\acute{c}\else \'{c}\fi{}, Angelova, Ansoldi, Antelis, Antier, Appert, Arai, Araya, Areeda, Ar\`ene, Arnaud, Aronson, Arun, Asali, Ashton, Assiduo, Aston, Astone, Aubin, Austin, Babak, Badaracco, Bader, Badger, Bae, Baer, Bagnasco, Bai, Baird, Ball, Ballardin, Ballmer, Balsamo, Baltus, Banagiri, Bankar, Barayoga, Barbieri, Barish, Barker, Barneo, Barone, Barr, Barsotti, Barsuglia, Barta, Bartlett, Barton, Bartos, Bassiri, Basti, Bawaj, Bayley, Baylor, Bazzan, B\'ecsy, Bedakihale, Bejger, Belahcene, Benedetto, Beniwal, Bennett, Bentley, BenYaala, Bergamin, Berger, Bernuzzi, Berry, Bersanetti, Bertolini, Betzwieser, Beveridge, Bhandare, Bhardwaj, Bhattacharjee, Bhaumik, Bilenko, Billingsley, Bini, Birney, Birnholtz, Biscans, Bischi,
  Biscoveanu, Bisht, Biswas, Bitossi, Bizouard, Blackburn, Blair, Blair, Blair, Bobba, Bode, Boer, Bogaert, Boldrini, Bonavena, Bondu, Bonilla, Bonnand, Booker, Boom, Bork, Boschi, Bose, Bose, Bossilkov, Boudart, Bouffanais, Bozzi, Bradaschia, Brady, Bramley, Branch, Branchesi, Brau, Breschi, Briant, Briggs, Brillet, Brinkmann, Brockill, Brooks, Brooks, Brown, Brunett, Bruno, Bruntz, Bryant, Bulik, Bulten, Buonanno, Buscicchio, Buskulic, Buy, Byer, Cadonati, Cagnoli, Cahillane, Bustillo, Callaghan, Callister, Calloni, Cameron, Camp, Canepa, Canevarolo, Cannavacciuolo, Cannon, Cao, Capote, Carapella, Carbognani, Carlin, Carney, Carpinelli, Carrillo, Carullo, Carver, Diaz, Casentini, Castaldi, Caudill, Cavagli\`a, Cavalier, Cavalieri, Ceasar, Cella, Cerd\'a-Dur\'an, Cesarini, Chaibi, Chakravarti, Subrahmanya, Champion, Chan, Chan, Chan, Chan, Chandra, Chanial, Chao, Charlton, Chase, Chassande-Mottin, Chatterjee, Chatterjee, Chatterjee, Chattopadhyay, Chaturvedi, Chaty, Chatziioannou, Chen, Chen, Chen, Chen,
  Chen, Cheng, Cheong, Cheung, Chia, Chiadini, Chiarini, Chierici, Chincarini, Chiofalo, Chiummo, Cho, Cho, Choudhary, Choudhary, Christensen, Chu, Chua, Chung, Ciani, Ciecielag, Cie\ifmmode~\acute{s}\else \'{s}\fi{}lar, Cifaldi, Ciobanu, Ciolfi, Cipriano, Cirone, Clara, Clark, Clark, Clarke, Clearwater, Clesse, Cleva, Coccia, Codazzo, Cohadon, Cohen, Cohen, Colleoni, Collette, Colombo, Colpi, Compton, Constancio, Conti, Cooper, Corban, Corbitt, Cordero-Carri\'on, Corezzi, Corley, Cornish, Corre, Corsi, Cortese, Costa, Cotesta, Coughlin, Coulon, Countryman, Cousins, Couvares, Coward, Cowart, Coyne, Coyne, Creighton, Creighton, Criswell, Croquette, Crowder, Cudell, Cullen, Cumming, Cummings, Cunningham, Cuoco, Cury\l{}o, Dabadie, Canton, Dall'Osso, D\'alya, Dana, DaneshgaranBajastani, D'Angelo, Danila, Danilishin, D'Antonio, Danzmann, Darsow-Fromm, Dasgupta, Datrier, Datta, Dattilo, Dave, Davier, Davies, Davis, Davis, Daw, Dean, DeBra, Deenadayalan, Degallaix, De~Laurentis, Del\'eglise, Del~Favero, De~Lillo,
  De~Lillo, Del~Pozzo, DeMarchi, De~Matteis, D'Emilio, Demos, Dent, Depasse, De~Pietri, De~Rosa, De~Rossi, DeSalvo, De~Simone, Dhurandhar, D\'{\i}az, Diaz-Ortiz, Didio, Dietrich, Di~Fiore, Di~Fronzo, Di~Giorgio, Di~Giovanni, Di~Giovanni, Di~Girolamo, Di~Lieto, Ding, Di~Pace, Di~Palma, Di~Renzo, Divakarla, Divyajyoti, Dmitriev, Doctor, D'Onofrio, Donovan, Dooley, Doravari, Dorrington, Drago, Driggers, Drori, Ducoin, Dupej, Durante, D'Urso, Duverne, Dwyer, Eassa, Easter, Ebersold, Eckhardt, Eddolls, Edelman, Edo, Edy, Effler, Eichholz, Eikenberry, Eisenmann, Eisenstein, Ejlli, Engelby, Errico, Essick, Estell\'es, Estevez, Etienne, Etzel, Evans, Evans, Ewing, Fafone, Fair, Fairhurst, Fanning, Farah, Farinon, Farr, Farr, Farrow, Fauchon-Jones, Favaro, Favata, Fays, Fazio, Feicht, Fejer, Fenyvesi, Ferguson, Fernandez-Galiana, Ferrante, Ferreira, Fidecaro, Figura, Fiori, Fishbach, Fisher, Fittipaldi, Fiumara, Flaminio, Floden, Fong, Font, Fornal, Forsyth, Franke, Frasca, Frasconi, Frederick, Freed, Frei, Freise,
  Frey, Fritschel, Frolov, Fronz\'e, Fulda, Fyffe, Gabbard, Gabella, Gadre, Gair, Gais, Galaudage, Gamba, Ganapathy, Ganguly, Gaonkar, Garaventa, Garc\'{\i}a, Garc\'{\i}a-N\'u\~nez, Garc\'{\i}a-Quir\'os, Garufi, Gateley, Gaudio, Gayathri, Gemme, Gennai, George, George, Gerberding, Gergely, Gewecke, Ghonge, Ghosh, Ghosh, Ghosh, Ghosh, Giacomazzo, Giacoppo, Giaime, Giardina, Gibson, Gier, Giesler, Giri, Gissi, Glanzer, Gleckl, Godwin, Goetz, Goetz, Gohlke, Goncharov, Gonz\'alez, Gopakumar, Gosselin, Gouaty, Gould, Grace, Grado, Granata, Granata, Grant, Gras, Grassia, Gray, Gray, Greco, Green, Green, Gretarsson, Gretarsson, Griffith, Griffiths, Griggs, Grignani, Grimaldi, Grimm, Grote, Grunewald, Gruning, Guerra, Guidi, Guimaraes, Guix\'e, Gulati, Guo, Guo, Gupta, Gupta, Gupta, Gustafson, Gustafson, Guzman, Haegel, Halim, Hall, Hamilton, Hammond, Haney, Hanks, Hanna, Hannam, Hannuksela, Hansen, Hansen, Hanson, Harder, Hardwick, Haris, Harms, Harry, Harry, Hartwig, Haskell, Hasskew, Haster, Haughian, Hayes,
  Healy, Heidmann, Heidt, Heintze, Heinze, Heinzel, Heitmann, Hellman, Hello, Helmling-Cornell, Hemming, Hendry, Heng, Hennes, Hennig, Hennig, Hernandez, Vivanco, Heurs, Hild, Hill, Hines, Hochheim, Hofman, Hohmann, Holcomb, Holland, Holley-Bockelmann, Hollows, Holmes, Holt, Holz, Hopkins, Hough, Hourihane, Howell, Hoy, Hoyland, Hreibi, Hsu, Huang, H\"ubner, Huddart, Hughey, Hui, Husa, Huttner, Huxford, Huynh-Dinh, Idzkowski, Iess, Ingram, Isi, Isleif, Iyer, JaberianHamedan, Jacqmin, Jadhav, Jadhav, James, Jan, Jani, Janquart, Janssens, Janthalur, Jaranowski, Jariwala, Jaume, Jenkins, Jenner, Jeunon, Jia, Johns, Johnson-McDaniel, Jones, Jones, Jones, Jones, Jones, Jonker, Ju, Junker, Juste, Kalaghatgi, Kalogera, Kamai, Kandhasamy, Kang, Kanner, Kao, Kapadia, Kapasi, Karat, Karathanasis, Karki, Kashyap, Kasprzack, Kastaun, Katsanevas, Katsavounidis, Katzman, Kaur, Kawabe, K\'ef\'elian, Keitel, Key, Khadka, Khalili, Khan, Khazanov, Khetan, Khursheed, Kijbunchoo, Kim, Kim, Kim, Kim, Kim, Kimball, Kinley-Hanlon,
  Kirchhoff, Kissel, Kleybolte, Klimenko, Knee, Knowles, Knyazev, Koch, Koekoek, Koley, Kolitsidou, Kolstein, Komori, Kondrashov, Kontos, Koper, Korobko, Kovalam, Kozak, Kringel, Krishnendu, Kr\'olak, Kuehn, Kuei, Kuijer, Kulkarni, Kumar, Kumar, Kumar, Kumar, Kuns, Kuwahara, Lagabbe, Laghi, Lalande, Lam, Lamberts, Landry, Lane, Lang, Lange, Lantz, La~Rosa, Lartaux-Vollard, Lasky, Laxen, Lazzarini, Lazzaro, Leaci, Leavey, Lecoeuche, Lee, Lee, Lee, Lee, Lehmann, Lema\^{\i}tre, Leroy, Letendre, Levesque, Levin, Leviton, Leyde, Li, Li, Li, Li, Li, Linde, Linker, Linley, Littenberg, Liu, Liu, Liu, Llamas, Llorens-Monteagudo, Lo, Lockwood, London, Longo, Lopez, Portilla, Lorenzini, Loriette, Lormand, Losurdo, Lott, Lough, Lousto, Lovelace, Lucaccioni, L\"uck, Lumaca, Lundgren, Lynam, Macas, MacInnis, Macleod, MacMillan, Macquet, Hernandez, Magazz\`u, Magee, Maggiore, Magnozzi, Mahesh, Majorana, Makarem, Maksimovic, Maliakal, Malik, Man, Mandic, Mangano, Mango, Mansell, Manske, Mantovani, Mapelli, Marchesoni,
  Marion, Mark, M\'arka, M\'arka, Markakis, Markosyan, Markowitz, Maros, Marquina, Marsat, Martelli, Martin, Martin, Martinez, Martinez, Martinez, Martinovic, Martynov, Marx, Masalehdan, Mason, Massera, Masserot, Massinger, Masso-Reid, Mastrogiovanni, Matas, Mateu-Lucena, Matichard, Matiushechkina, Mavalvala, McCann, McCarthy, McClelland, McClincy, McCormick, McCuller, McGhee, McGuire, McIsaac, McIver, McRae, McWilliams, Meacher, Mehmet, Mehta, Meijer, Melatos, Melchor, Mendell, Menendez-Vazquez, Menoni, Mercer, Mereni, Merfeld, Merilh, Merritt, Merzougui, Meshkov, Messenger, Messick, Meyers, Meylahn, Mhaske, Miani, Miao, Michaloliakos, Michel, Middleton, Milano, Miller, Miller, Miller, Millhouse, Mills, Milotti, Minazzoli, Minenkov, Mir, Miravet-Ten\'es, Mishra, Mishra, Mistry, Mitra, Mitrofanov, Mitselmakher, Mittleman, Mo, Moguel, Mogushi, Mohapatra, Mohite, Molina, Molina-Ruiz, Mondin, Montani, Moore, Moraru, Morawski, More, Moreno, Moreno, Morisaki, Mours, Mow-Lowry, Mozzon, Muciaccia, Mukherjee,
  Mukherjee, Mukherjee, Mukherjee, Mukherjee, Mukund, Mullavey, Munch, Mu\~niz, Murray, Musenich, Muusse, Nadji, Nagar, Napolano, Nardecchia, Naticchioni, Nayak, Nayak, Neil, Neilson, Nelemans, Nelson, Nery, Neubauer, Neunzert, Ng, Ng, Nguyen, Nguyen, Nguyen, Nichols, Nissanke, Nitoglia, Nocera, Norman, North, Nuttall, Oberling, O'Brien, O'Dell, Oelker, Oganesyan, Oh, Oh, Ohme, Ohta, Okada, Olivetto, Oram, O'Reilly, Ormiston, Ormsby, Ortega, O'Shaughnessy, O'Shea, Ossokine, Osthelder, Ottaway, Overmier, Pace, Pagano, Page, Pagliaroli, Pai, Pai, Palamos, Palashov, Palomba, Pan, Panda, Pang, Pankow, Pannarale, Pant, Panther, Paoletti, Paoli, Paolone, Park, Parker, Pascucci, Pasqualetti, Passaquieti, Passuello, Patel, Pathak, Patricelli, Patron, Patrone, Paul, Payne, Pedraza, Pegoraro, Pele, Penn, Perego, Pereira, Pereira, Perez, P\'erigois, Perkins, Perreca, Perri\`es, Petermann, Petterson, Pfeiffer, Pham, Phukon, Piccinni, Pichot, Piendibene, Piergiovanni, Pierini, Pierro, Pillant, Pillas, Pilo, Pinard, Pinto,
  Pinto, Piotrzkowski, Pirello, Pitkin, Placidi, Planas, Plastino, Pluchar, Poggiani, Polini, Pong, Ponrathnam, Popolizio, Porter, Poulton, Powell, Pracchia, Pradier, Prajapati, Prasai, Prasanna, Pratten, Principe, Prodi, Prokhorov, Prosposito, Prudenzi, Puecher, Punturo, Puosi, Puppo, P\"urrer, Qi, Quetschke, Quitzow-James, Raab, Raaijmakers, Radkins, Radulesco, Raffai, Rail, Raja, Rajan, Ramirez, Ramirez, Ramos-Buades, Rana, Rapagnani, Rapol, Ray, Raymond, Raza, Razzano, Read, Rees, Regimbau, Rei, Reid, Reid, Reitze, Relton, Renzini, Rettegno, Reza, Rezac, Ricci, Richards, Richardson, Richardson, Riemenschneider, Riles, Rinaldi, Rink, Rizzo, Robertson, Robie, Robinet, Rocchi, Rodriguez, Rolland, Rollins, Romanelli, Romano, Romel, Romero-Rodr\'{\i}guez, Romero-Shaw, Romie, Ronchini, Rosa, Rose, Rosell, Rosi\ifmmode~\acute{n}\else \'{n}\fi{}ska, Ross, Rowan, Rowlinson, Roy, Roy, Roy, Rozza, Ruggi, Ruiz-Rocha, Ryan, Sachdev, Sadecki, Sadiq, Sakellariadou, Salafia, Salconi, Saleem, Salemi, Samajdar, Sanchez,
  Sanchez, Sanchez, Sanchis-Gual, Sanders, Sanuy, Saravanan, Sarin, Sassolas, Satari, Sauter, Savage, Sawant, Sawant, Sayah, Schaetzl, Scheel, Scheuer, Schiworski, Schmidt, Schmidt, Schnabel, Schneewind, Schofield, Sch\"onbeck, Schulte, Schutz, Schwartz, Scott, Scott, Seglar-Arroyo, Sellers, Sengupta, Sentenac, Seo, Sequino, Sergeev, Setyawati, Shaffer, Shahriar, Shams, Sharma, Sharma, Shawhan, Shcheblanov, Shikauchi, Shoemaker, Shoemaker, ShyamSundar, Sieniawska, Sigg, Singer, Singh, Singh, Singha, Sintes, Sipala, Skliris, Slagmolen, Slaven-Blair, Smetana, Smith, Smith, Soldateschi, Somala, Son, Soni, Soni, Sordini, Sorrentino, Sorrentino, Soulard, Souradeep, Sowell, Spagnuolo, Spencer, Spera, Srinivasan, Srivastava, Srivastava, Staats, Stachie, Steer, Steinhoff, Steinlechner, Steinlechner, Stevenson, Stops, Stover, Strain, Strang, Stratta, Strunk, Sturani, Stuver, Sudhagar, Sudhir, Suh, Summerscales, Sun, Sun, Sunil, Sur, Suresh, Sutton, Swinkels, Szczepa\ifmmode~\acute{n}\else \'{n}\fi{}czyk, Szewczyk,
  Tacca, Tait, Talbot, Talbot, Tanasijczuk, Tanner, Tao, Tao, Mart\'{\i}n, Taranto, Tasson, Tenorio, Terhune, Terkowski, Thirugnanasambandam, Thomas, Thomas, Thomas, Thompson, Thondapu, Thorne, Thrane, Tiwari, Tiwari, Tiwari, Toivonen, Toland, Tolley, Tonelli, Torres-Forn\'e, Torrie, e~Melo, T\"oyr\"a, Trapananti, Travasso, Traylor, Trevor, Tringali, Tripathee, Troiano, Trovato, Trozzo, Trudeau, Tsai, Tsai, Tsang, Tse, Tso, Tsukada, Tsuna, Tsutsui, Turbang, Turconi, Ubhi, Udall, Ueno, Unnikrishnan, Urban, Utina, Vahlbruch, Vajente, Vajpeyi, Valdes, Valentini, Valsan, van Bakel, van Beuzekom, van~den Brand, Van Den~Broeck, Vander-Hyde, van~der Schaaf, van Heijningen, Vanosky, van Remortel, Vardaro, Vargas, Varma, Vas\'uth, Vecchio, Vedovato, Veitch, Veitch, Venneberg, Venugopalan, Verkindt, Verma, Verma, Veske, Vetrano, Vicer\'e, Vidyant, Viets, Vijaykumar, Villa-Ortega, Vinet, Virtuoso, Vitale, Vo, Vocca, von Reis, von Wrangel, Vorvick, Vyatchanin, Wade, Wade, Wagner, Walet, Walker, Wallace, Wallace, Walsh,
  Wang, Wang, Ward, Warner, Was, Washington, Watchi, Weaver, Webster, Weinert, Weinstein, Weiss, Weller, Weller, Wellmann, Wen, We\ss{}els, Wette, Whelan, White, Whiting, Whittle, Wilken, Williams, Williams, Williamson, Willis, Willke, Wilson, Winkler, Wipf, Wlodarczyk, Woan, Woehler, Wofford, Wong, Wu, Wysocki, Xiao, Yamamoto, Yang, Yang, Yang, Yang, Yap, Yeeles, Yelikar, Ying, Yoo, Yu, Yu, Zadro\ifmmode~\dot{z}\else \.{z}\fi{}ny, Zanolin, Zelenova, Zendri, Zevin, Zhang, Zhang, Zhang, Zhang, Zhao, Zhao, Zhao, Zhou, Zhou, Zhu, Zimmerman, Zlochower, Zucker, \& Zweizig}]{GWTC-2_1}
---. 2024, Phys. Rev. D, 109, 022001, \dodoi{10.1103/PhysRevD.109.022001}

\bibitem[{Acernese {et~al.}(2014)Acernese, Agathos, Agatsuma, Aisa, Allemandou, Allocca, Amarni, Astone, Balestri, Ballardin, Barone, Baronick, Barsuglia, Basti, Basti, Bauer, Bavigadda, Bejger, Beker, Belczynski, Bersanetti, Bertolini, Bitossi, \& Bizouard}]{Virgo}
Acernese, F., Agathos, M., Agatsuma, K., {et~al.} 2014, Classical and Quantum Gravity, 32, 024001, \dodoi{10.1088/0264-9381/32/2/024001}

\bibitem[{Bacon {et~al.}(2022)Bacon, Trovato, \& Bejger}]{Bacon_denoising}
Bacon, P., Trovato, A., \& Bejger, M. 2022, Denoising gravitational-wave signals from binary black holes with dilated convolutional autoencoder.
\newblock \doarXiv{2205.13513}

\bibitem[{Bahaadini {et~al.}(2017)Bahaadini, Rohani, Coughlin, Zevin, Kalogera, \& Katsaggelos}]{glitch3}
Bahaadini, S., Rohani, N., Coughlin, S., {et~al.} 2017, Deep Multi-view Models for Glitch Classification.
\newblock \doarXiv{1705.00034}

\bibitem[{Berti {et~al.}(2018)Berti, Yagi, \& Yunes}]{TGR}
Berti, E., Yagi, K., \& Yunes, N. 2018, Gen. Rel. Grav., 50, 46, \dodoi{10.1007/s10714-018-2362-8}

\bibitem[{Capote {et~al.}(2024)Capote, Dartez, \& Davis}]{Glitch_overlapping_with_signal}
Capote, E., Dartez, L., \& Davis, D. 2024, Technical Noise, Data Quality, and Calibration Requirements for Next-Generation Gravitational-Wave Science.
\newblock \doarXiv{2404.04761}

\bibitem[{Chatterjee \& Jani(2024{\natexlab{a}})}]{AWaRe}
Chatterjee, C., \& Jani, K. 2024{\natexlab{a}}, The Astrophysical Journal, 969, 25, \dodoi{10.3847/1538-4357/ad4602}

\bibitem[{Chatterjee \& Jani(2024{\natexlab{b}})}]{AWaRe_uncertainty}
---. 2024{\natexlab{b}}, The Astrophysical Journal, 973, 112, \dodoi{10.3847/1538-4357/ad6984}

\bibitem[{Chatterjee {et~al.}(2021)Chatterjee, Wen, Diakogiannis, \& Vinsen}]{Chatterjee_2021}
Chatterjee, C., Wen, L., Diakogiannis, F., \& Vinsen, K. 2021, Phys. Rev. D, 104, 064046, \dodoi{10.1103/PhysRevD.104.064046}

\bibitem[{Collaboration {et~al.}(2023)Collaboration, the Virgo~Collaboration, the KAGRA~Collaboration, Abbott, Abe, Acernese, Ackley, Adhicary, Adhikari, Adhikari, Adkins, Adya, Affeldt, Agarwal, Agathos, Aguiar, Aiello, Ain, Ajith, Akutsu, Albanesi, Alfaidi, Al-Jodah, Alléné, \& Allocca}]{GWOSC}
Collaboration, T. L.~S., the Virgo~Collaboration, the KAGRA~Collaboration, {et~al.} 2023, Open data from the third observing run of LIGO, Virgo, KAGRA and GEO.
\newblock \doarXiv{2302.03676}

\bibitem[{Cornish \& Littenberg(2015)}]{BayesWave}
Cornish, N.~J., \& Littenberg, T.~B. 2015, Classical and Quantum Gravity, 32, 135012, \dodoi{10.1088/0264-9381/32/13/135012}

\bibitem[{Davis {et~al.}(2022)Davis, Littenberg, Romero-Shaw, Millhouse, McIver, Renzo, \& Ashton}]{O3_glitch_subtraction}
Davis, D., Littenberg, T.~B., Romero-Shaw, I.~M., {et~al.} 2022, Classical and Quantum Gravity, 39, 245013, \dodoi{10.1088/1361-6382/aca238}

\bibitem[{Davis {et~al.}(2021)Davis, Areeda, Berger, Bruntz, Effler, Essick, Fisher, Godwin, Goetz, Helmling-Cornell, Hughey, Katsavounidis, Lundgren, Macleod, Márka, Massinger, Matas, McIver, Mo, Mogushi, Nguyen, Nuttall, Schofield, Shoemaker, Soni, Stuver, Urban, Valdes, Walker, Abbott, Adams, Adhikari, Ananyeva, Appert, Arai, Asali, Aston, Austin, Baer, Ball, Ballmer, Banagiri, Barker, Barschaw, Barsotti, Bartlett, Betzwieser, Beda, Bhattacharjee, Bidler, Billingsley, Biscans, Blair, Blair, Bode, Booker, Bork, Bramley, Brooks, Brown, Buikema, Cahillane, Callister, Santoro, Cannon, Carlin, Chandra, Chen, Christensen, Ciobanu, Clara, Compton, Cooper, Corley, Coughlin, Countryman, Covas, Coyne, Crowder, Canton, Danila, Datrier, Davies, Dent, Didio, Fronzo, Dooley, Driggers, Dupej, Dwyer, Etzel, Evans, Evans, Fairhurst, Feicht, Fernandez-Galiana, Frey, Fritschel, Frolov, Fulda, Fyffe, Gadre, Giaime, Giardina, González, Gras, Gray, Gray, Green, Gupta, Gustafson, Gustafson, Hanks, Hanson, Hardwick, Harry,
  Hasskew, Heintze, Heinzel, Holland, Hollows, Hoy, Hughey, Jadhav, Janssens, Johns, Jones, Kandhasamy, Karki, Kasprzack, Kawabe, Keitel, Kijbunchoo, Kim, King, Kissel, Kulkarni, Kumar, Landry, Lane, Lantz, Laxen, Lecoeuche, Leviton, Liu, Lormand, Macas, Macedo, MacInnis, Mandic, Mansell, Márka, Martinez, Martinovic, Martynov, Mason, Matichard, Mavalvala, McCarthy, McClelland, McCormick, McCuller, McIsaac, McRae, Mendell, Merfeld, Merilh, Meyers, Meylahn, Michaloliakos, Middleton, Mills, Mistry, Mittleman, Moreno, Mow-Lowry, Mozzon, Mueller, Mukund, Mullavey, Muth, Nelson, Neunzert, Nichols, Nitoglia, Oberling, Oh, Oh, Oram, Ormiston, Ormsby, Osthelder, Ottaway, Overmier, Pai, Palamos, Pannarale, Parker, Patane, Patel, Payne, Pele, Penhorwood, Perez, Phukon, Pillas, Pirello, Radkins, Ramirez, Richardson, Riles, Rink, Robertson, Rollins, Romel, Romie, Ross, Ryan, Sadecki, Sakellariadou, Sanchez, Sanchez, Sandles, Saravanan, Savage, Schaetzl, Schnabel, Schwartz, Sellers, Shaffer, Sigg, Sintes, Slagmolen,
  Smith, Soni, Sorazu, Spencer, Strain, Strom, Sun, Szczepańczyk, Tasson, Tenorio, Thomas, Thomas, Thorne, Toland, Torrie, Tran, Traylor, Trevor, Tse, Vajente, van Remortel, Vander-Hyde, Vargas, Veitch, Veitch, Venkateswara, Venugopalan, Viets, Villa-Ortega, Vo, Vorvick, Wade, Wallace, Ward, Warner, Weaver, Weinstein, Weiss, Wette, White, White, Whittle, Williamson, Willke, Wipf, Xiao, Xu, Yamamoto, Yu, Yu, Zhang, Zheng, Zucker, \& Zweizig}]{glitches_definition3}
Davis, D., Areeda, J.~S., Berger, B.~K., {et~al.} 2021, Classical and Quantum Gravity, 38, 135014, \dodoi{10.1088/1361-6382/abfd85}

\bibitem[{George {et~al.}(2018)George, Shen, \& Huerta}]{glitch2}
George, D., Shen, H., \& Huerta, E.~A. 2018, Phys. Rev. D, 97, 101501, \dodoi{10.1103/PhysRevD.97.101501}

\bibitem[{Ghonge {et~al.}(2024)Ghonge, Brandt, Sullivan, Millhouse, Chatziioannou, Clark, Littenberg, Cornish, Hourihane, \& Cadonati}]{Impact_of_glitch6}
Ghonge, S., Brandt, J., Sullivan, J.~M., {et~al.} 2024, Assessing and Mitigating the Impact of Glitches on Gravitational-Wave Parameter Estimation: a Model Agnostic Approach.
\newblock \doarXiv{2311.09159}

\bibitem[{Glanzer {et~al.}(2023)Glanzer, Banagiri, Coughlin, Soni, Zevin, Berry, Patane, Bahaadini, Rohani, Crowston, Kalogera, Østerlund, Trouille, \& Katsaggelos}]{GravitySpy_O3}
Glanzer, J., Banagiri, S., Coughlin, S.~B., {et~al.} 2023, Classical and Quantum Gravity, 40, 065004, \dodoi{10.1088/1361-6382/acb633}

\bibitem[{{Hannam} {et~al.}(2022){Hannam}, {Hoy}, {Thompson}, {Fairhurst}, {Raymond}, {Colleoni}, {Davis}, {Estell{\'e}s}, {Haster}, {Helmling-Cornell}, {Husa}, {Keitel}, {Massinger}, {Men{\'e}ndez-V{\'a}zquez}, {Mogushi}, {Ossokine}, {Payne}, {Pratten}, {Romero-Shaw}, {Sadiq}, {Schmidt}, {Tenorio}, {Udall}, {Veitch}, {Williams}, {Yelikar}, \& {Zimmerman}}]{GW200129_1}
{Hannam}, M., {Hoy}, C., {Thompson}, J.~E., {et~al.} 2022, \nat, 610, 652, \dodoi{10.1038/s41586-022-05212-z}

\bibitem[{Hochreiter \& Schmidhuber(1997)}]{LSTM}
Hochreiter, S., \& Schmidhuber, J. 1997, Neural Computation, 9, 1735

\bibitem[{Hourihane {et~al.}(2022)Hourihane, Chatziioannou, Wijngaarden, Davis, Littenberg, \& Cornish}]{Impact_of_glitch5}
Hourihane, S., Chatziioannou, K., Wijngaarden, M., {et~al.} 2022, Phys. Rev. D, 106, 042006, \dodoi{10.1103/PhysRevD.106.042006}

\bibitem[{Klimenko {et~al.}(2016)Klimenko, Vedovato, Drago, Salemi, Tiwari, Prodi, Lazzaro, Ackley, Tiwari, Da~Silva, \& Mitselmakher}]{cWB}
Klimenko, S., Vedovato, G., Drago, M., {et~al.} 2016, Phys. Rev. D, 93, 042004, \dodoi{10.1103/PhysRevD.93.042004}

\bibitem[{Krizhevsky {et~al.}(2012)Krizhevsky, Sutskever, \& Hinton}]{CNN_2}
Krizhevsky, A., Sutskever, I., \& Hinton, G.~E. 2012, in Advances in Neural Information Processing Systems, ed. F.~Pereira, C.~Burges, L.~Bottou, \& K.~Weinberger, Vol.~25 (Curran Associates, Inc.)

\bibitem[{Kwok {et~al.}(2022)Kwok, Lo, Weinstein, \& Li}]{Impact_of_glitch2}
Kwok, J. Y.~L., Lo, R. K.~L., Weinstein, A.~J., \& Li, T. G.~F. 2022, Phys. Rev. D, 105, 024066, \dodoi{10.1103/PhysRevD.105.024066}

\bibitem[{Lecun {et~al.}(1998)Lecun, Bottou, Bengio, \& Haffner}]{CNN_1}
Lecun, Y., Bottou, L., Bengio, Y., \& Haffner, P. 1998, Proceedings of the IEEE, 86, 2278, \dodoi{10.1109/5.726791}

\bibitem[{Llorens-Monteagudo {et~al.}(2019)Llorens-Monteagudo, Torres-Forné, Font, \& Marquina}]{glitch5}
Llorens-Monteagudo, M., Torres-Forné, A., Font, J.~A., \& Marquina, A. 2019, Classical and Quantum Gravity, 36, 075005, \dodoi{10.1088/1361-6382/ab0657}

\bibitem[{Macas {et~al.}(2024)Macas, Lundgren, \& Ashton}]{GW200129_3}
Macas, R., Lundgren, A., \& Ashton, G. 2024, Phys. Rev. D, 109, 062006, \dodoi{10.1103/PhysRevD.109.062006}

\bibitem[{Macas {et~al.}(2022)Macas, Pooley, Nuttall, Davis, Dyer, Lecoeuche, Lyman, McIver, \& Rink}]{Impact_of_glitch4}
Macas, R., Pooley, J., Nuttall, L.~K., {et~al.} 2022, Phys. Rev. D, 105, 103021, \dodoi{10.1103/PhysRevD.105.103021}

\bibitem[{{Margalit} \& {Metzger}(2017)}]{GW170817_EOS}
{Margalit}, B., \& {Metzger}, B.~D. 2017, The Astrophysical Journal Letters, 850, L19, \dodoi{10.3847/2041-8213/aa991c}

\bibitem[{Mozzon {et~al.}(2022)Mozzon, Ashton, Nuttall, \& Williamson}]{Impact_of_glitch3}
Mozzon, S., Ashton, G., Nuttall, L.~K., \& Williamson, A.~R. 2022, Physical Review D, 106, \dodoi{10.1103/physrevd.106.043504}

\bibitem[{Pankow {et~al.}(2018)Pankow, Chatziioannou, Chase, Littenberg, Evans, McIver, Cornish, Haster, Kanner, Raymond, Vitale, \& Zimmerman}]{GW170817_glitch}
Pankow, C., Chatziioannou, K., Chase, E.~A., {et~al.} 2018, Phys. Rev. D, 98, 084016, \dodoi{10.1103/PhysRevD.98.084016}

\bibitem[{Payne {et~al.}(2022)Payne, Hourihane, Golomb, Udall, Davis, \& Chatziioannou}]{GW200129_4}
Payne, E., Hourihane, S., Golomb, J., {et~al.} 2022, Phys. Rev. D, 106, 104017, \dodoi{10.1103/PhysRevD.106.104017}

\bibitem[{Powell(2018)}]{Impact_of_glitch1}
Powell, J. 2018, Classical and Quantum Gravity, 35, 155017, \dodoi{10.1088/1361-6382/aacf18}

\bibitem[{Powell {et~al.}(2017)Powell, Torres-Forné, Lynch, Trifirò, Cuoco, Cavaglià, Heng, \& Font}]{glitch4}
Powell, J., Torres-Forné, A., Lynch, R., {et~al.} 2017, Classical and Quantum Gravity, 34, 034002, \dodoi{10.1088/1361-6382/34/3/034002}

\bibitem[{Pratten {et~al.}(2021)Pratten, Garc\'{\i}a-Quir\'os, Colleoni, Ramos-Buades, Estell\'es, Mateu-Lucena, Jaume, Haney, Keitel, Thompson, \& Husa}]{IMRPhenomXPHM}
Pratten, G., Garc\'{\i}a-Quir\'os, C., Colleoni, M., {et~al.} 2021, Phys. Rev. D, 103, 104056, \dodoi{10.1103/PhysRevD.103.104056}

\bibitem[{Razzano \& Cuoco(2018)}]{glitch6}
Razzano, M., \& Cuoco, E. 2018, Classical and Quantum Gravity, 35, 095016, \dodoi{10.1088/1361-6382/aab793}

\bibitem[{Selvaraju {et~al.}(2020)Selvaraju, Cogswell, Das, Vedantam, Parikh, \& Batra}]{Grad-CAM}
Selvaraju, R.~R., Cogswell, M., Das, A., {et~al.} 2020, International Journal of Computer Vision, 128, 336, \dodoi{10.1007/s11263-019-01228-7}

\bibitem[{Soni {et~al.}(2021)Soni, Berry, Coughlin, Harandi, Jackson, Crowston, Østerlund, Patane, Katsaggelos, Trouille, Baranowski, Domainko, Kaminski, Rodriguez, Marciniak, Nauta, Niklasch, Rote, Téglás, Unsworth, \& Zhang}]{GravitySpy_classifier}
Soni, S., Berry, C. P.~L., Coughlin, S.~B., {et~al.} 2021, Classical and Quantum Gravity, 38, 195016, \dodoi{10.1088/1361-6382/ac1ccb}

\bibitem[{Soni {et~al.}(2024)Soni, Berger, Davis, Renzo, Effler, Ferreira, Glanzer, Goetz, González, Helmling-Cornell, Hughey, Huxford, Mannix, Mo, Nandi, Neunzert, Nichols, Pham, Renzini, Schofield, Stuver, Trevor, Álvarez López, Beda, Berry, Bhuiyan, Bruntz, Christensen, Blagg, Chan, Charlton, Connolly, Dhatri, Ding, Garg, Holley-Bockelmann, Hourihane, Jani, Janssens, Jarov, Knee, Lattal, Lecoeuche, Littenberg, Liyanage, Lott, Macas, Malakar, McGowan, McIver, Millhouse, Nuttall, Nykamp, Ota, Rawcliffe, Scully, Tasson, Tejera, Thiele, Udall, Winborn, Yarbrough, Zhang, Abbott, Abouelfettouh, Adhikari, Ananyeva, Appert, Arai, Aritomi, Aston, Ball, Ballmer, Barker, Barsotti, Betzwieser, Billingsley, Biscans, Bode, Bonilla, Bossilkov, Branch, Brooks, Brown, Bryant, Cahillane, Cao, Capote, Clara, Collins, Compton, Cottingham, Coyne, Crouch, Csizmazia, Cullen, Dartez, Demos, Dohmen, Driggers, Dwyer, Ejlli, Etzel, Evans, Feicht, Frey, Frischhertz, Fritschel, Frolov, Fulda, Fyffe, Ganapathy, Gateley, Giaime,
  Giardina, Goetz, Goodwin-Jones, Gras, Gray, Griffith, Grote, Guidry, Hall, Hanks, Hanson, Heintze, Holland, Hoyland, Huang, Inoue, James, Jennings, Jia, Karat, Karki, Kasprzack, Kawabe, Kijbunchoo, King, Kissel, Komori, Kontos, Kumar, Kuns, Landry, Lantz, Laxen, Lee, Lesovsky, Llamas, Lormand, Loughlin, Macas, MacInnis, Makarem, Mansell, Martin, Mason, Matichard, Mavalvala, Maxwell, McCarrol, McCarthy, McClelland, McCormick, McCuller, McRae, Mera, Merilh, Meylahn, Mittleman, Moraru, Moreno, Mullavey, Nakano, Nelson, Notte, Oberling, O'Hanlon, Osthelder, Ottaway, Overmier, Parker, Pele, Pham, Pirello, Quetschke, Ramirez, Reyes, Richardson, Robinson, Rollins, Romel, Romie, Ross, Ryan, Sadecki, Sanchez, Sanchez, Sanchez, Savage, Schaetzl, Schiworski, Schnabel, Schwartz, Sellers, Shaffer, Short, Sigg, Slagmolen, Soike, Srivastava, Sun, Tanner, Thomas, Thomas, Thorne, Torrie, Traylor, Ubhi, Vajente, Vanosky, Vecchio, Veitch, Vibhute, von Reis, Warner, Weaver, Weiss, Whittle, Willke, Wipf, Xu, Yamamoto, Zhang, \&
  Zucker}]{glitches_definition2}
Soni, S., Berger, B.~K., Davis, D., {et~al.} 2024, LIGO Detector Characterization in the first half of the fourth Observing run.
\newblock \doarXiv{2409.02831}

\bibitem[{Sánchez {et~al.}(2019)Sánchez, {Domínguez R.}, Lares, Beroiz, Cabral, Gurovich, Quiñones, Artola, Colazo, Schneiter, Girardini, Tornatore, {Nilo Castellón}, {García Lambas}, \& Díaz}]{glitch1}
Sánchez, B., {Domínguez R.}, M., Lares, M., {et~al.} 2019, Astronomy and Computing, 28, 100284, \dodoi{https://doi.org/10.1016/j.ascom.2019.05.002}

\bibitem[{Udall {et~al.}(2024)Udall, Hourihane, Miller, Davis, Chatziioannou, Isi, \& Deshong}]{GW191109_2}
Udall, R., Hourihane, S., Miller, S., {et~al.} 2024, The anti-aligned spin of GW191109: glitch mitigation and its implications.
\newblock \doarXiv{2409.03912}

\bibitem[{Varma {et~al.}(2022)Varma, Biscoveanu, Islam, Shaik, Haster, Isi, Farr, Field, \& Vitale}]{GW200129_2}
Varma, V., Biscoveanu, S., Islam, T., {et~al.} 2022, Phys. Rev. Lett., 128, 191102, \dodoi{10.1103/PhysRevLett.128.191102}

\bibitem[{Vaswani {et~al.}(2017)Vaswani, Shazeer, Parmar, Uszkoreit, Jones, Gomez, Kaiser, \& Polosukhin}]{Transformer}
Vaswani, A., Shazeer, N., Parmar, N., {et~al.} 2017, CoRR, abs/1706.03762

\bibitem[{Zevin {et~al.}(2017)Zevin, Coughlin, Bahaadini, Besler, Rohani, Allen, Cabero, Crowston, Katsaggelos, Larson, Lee, Lintott, Littenberg, Lundgren, Østerlund, Smith, Trouille, \& Kalogera}]{GravitySpy}
Zevin, M., Coughlin, S., Bahaadini, S., {et~al.} 2017, Classical and Quantum Gravity, 34, 064003, \dodoi{10.1088/1361-6382/aa5cea}

\bibitem[{Zevin {et~al.}(2024)}]{GravitySpy1}
Zevin, M., {et~al.} 2024, Eur. Phys. J. Plus, 139, 100, \dodoi{10.1140/epjp/s13360-023-04795-4}

\bibitem[{Zhang {et~al.}(2023)Zhang, Fragione, Kimball, \& Kalogera}]{GW191109_1}
Zhang, R.~C., Fragione, G., Kimball, C., \& Kalogera, V. 2023, The Astrophysical Journal, 954, 23, \dodoi{10.3847/1538-4357/ace4c1}

\end{thebibliography}
\bibliographystyle{aasjournal}



\end{document}